\documentclass[twocolumn,amssymb,nobibnotes,aps,prx,superscriptaddress]{revtex4-2}

\usepackage{graphicx}
\usepackage{amsmath, amsthm, amssymb}
\theoremstyle{definition}

\usepackage{mathdots}
\usepackage{float}
\usepackage{pgfplots} 
\usepackage{tikz-3dplot}
\usepgfplotslibrary{patchplots}
\usepgfplotslibrary{groupplots}
\pgfplotsset{compat=1.6,ylabsh/.style={every axis y label/.style={at={(0,0.5)}, xshift=#1, rotate=90}}}
\usetikzlibrary{positioning}
\usetikzlibrary{shapes}
\usetikzlibrary{shapes.geometric}
\usetikzlibrary{decorations.text}
\usetikzlibrary{decorations.pathreplacing,shapes.misc}
\usetikzlibrary{matrix}
\usepackage{caption,subcaption}
\usepackage[ruled,lined]{algorithm2e}
\usepackage{algorithmic}
\renewcommand{\algorithmiccomment}[1]{\bgroup\hfill//~#1\egroup}

\newenvironment{frcseries}{\fontfamily{frc}\selectfont}{}

\begin{document}

\newcommand{\papertitle}{Collective Cell Fluidity Controls Active Prestress Transmission in Cell–Extracellular-Matrix Tissues}
\title{\papertitle}
\author{Liyang Wang}
\affiliation{School of Chemistry and Chemical Engineering, Shanghai Jiao Tong University, Shanghai 200240, China}
\author{J. M. Schwarz}
\email{jmschw02@syr.edu}
\affiliation{Physics Department, Syracuse University, Syracuse, NY 13244, USA}
\affiliation{Indian Creek Farm, Ithaca, NY 14850, USA}
\author{Tao Zhang}
\email{zhangtao.scholar@sjtu.edu.cn}
\affiliation{School of Chemistry and Chemical Engineering, Shanghai Jiao Tong University, Shanghai 200240, China}

\date{\today}
\begin{abstract}
Tissues are active composite materials in which multicellular collectives and extracellular matrices continuously exert forces on and mechanically reorganize one another. Understanding tissue mechanics therefore requires resolving not only the mechanical properties of cells and the extracellular matrix (ECM), but also the dynamic interface through which forces are transmitted between them. Here, we develop a three-dimensional micromechanical model of a cell–ECM tissue that explicitly couples deformable, rearranging cell clusters to a disordered network of semiflexible fibers through a dynamic, force-generating interface. Cell clusters are represented as multicellular spheroids described by a three-dimensional vertex model, with solid-like or fluid-like collective dynamics, and are coupled to the ECM by passive or actively contractile linkers that are renewed as the cell-cluster boundary reorganizes. Comparisons with matched intact, voided, and passive-linker networks separate cavity formation, passive interfacial tethering, and active loading. At small strain, passive tethering provides only modest reinforcement, whereas active contraction prestresses the matrix and recruits tensile load-bearing pathways. Solid-like cell clusters preserve coherent load paths and exhibit an excess modulus that scales approximately as \(|\sigma|^{1.4}\) across variations in activity, cluster size, and cluster number. Fluid-like cell clusters undergo greater interfacial renewal, producing weaker and nonmonotonic coupling between prestress and stiffness. Increasing cell-cluster number reveals collective stiffening when prestressed regions become connected through sufficiently persistent cell–ECM interfaces. At large strain, both solid-like and fluid-like systems approach the corresponding voided-network response as the residual fiber backbone becomes mechanically dominant. These results establish a micromechanical principle for active cell–ECM tissues: macroscopic mechanics is controlled not only by the magnitude of cell-generated prestress, but by the ability of collective cell dynamics and the cell–ECM interface to organize and persistently transmit that prestress through the extracellular matrix.
\end{abstract}
\maketitle
		
\section{Introduction}

Tissues are active composite materials in which collectively rearranging cells generate forces that are transmitted through a deformable extracellular matrix (ECM)~\cite{mouw2014extracellular,anlacs2018tissue,humphrey2014mechanotransduction}. Their macroscopic mechanics therefore depends not only on the material properties of cells and ECM, but also on how forces are generated and transmitted across the dynamic interface between them~\cite{humphrey2014mechanotransduction,naba2024mechanisms,stramer2024extracellular}. Connecting these microscopic processes to tissue-scale mechanics requires models that explicitly resolve both cellular and extracellular degrees of freedom. In particular, how the collective material state of cells regulates the transmission of actively generated stresses through the ECM remains poorly understood. Here, we develop a three-dimensional micromechanical model of a cell–ECM tissue consisting of multicellular spheroids (or clusters) embedded in fibrous extracellular matrices (ECMs) to form an active composite in which cell--cell mechanics, cell--matrix coupling, and matrix deformation are strongly intertwined. The ECM regulates collective cell morphology, migration, and invasion, while cellular contraction reorganizes and mechanically loads the surrounding fibers. 

For single spheroids embedded in collagen, experiments have shown that tumor spheroids generate long-ranged deformations in collagen networks and that their collective forces evolve through mechanical feedback with the ECM~\cite{Mark2020CollectiveForces}. Complementary osmotic-compression experiments and active-poroelastic modeling showed that preexisting active stress modifies the emergent bulk modulus and hydraulic diffusion of multicellular spheroids~\cite{Dolega2021}. At the single-cell scale, contraction can produce extended stress and stiffness gradients in collagen, fibrin, and Matrigel because fibrous architectures transmit tensile forces over distances much larger than the cell size~\cite{Hall2016,Han2018StressStiffening}. An embedded spheroid is therefore not simply a passive inclusion, but a deformable active body that loads the matrix through a distributed and evolving interface~\cite{zhang2025}. The central mechanical question is not only how much prestress a spheroid generates, but how efficiently that prestress is converted into macroscopic shear stiffness when the interface transmitting it can rearrange.

This distinction is particularly important in collagen-like networks, whose connectivity is often below the central-force isostatic threshold. Finite fiber-bending elasticity stabilizes otherwise floppy modes, producing a compliant and strongly nonaffine response at small strain, whereas increasing deformation recruits tensile pathways and drives nonlinear stiffening~\cite{Mao2010, das2012redundancy,Sharma2016-kf,    Heidemann2015Elasticity3DNetworks}. Internal motor stresses can stabilize underconnected networks by suppressing nonaffine fluctuations~\cite{Sheinman2012PRL}, while externally imposed or boundary-generated prestress can shift the onset of nonlinear stiffening~\cite{Vahabi2016AxialPrestress,Arzash2019}. In bucklable networks, contractile forcing is further rectified: compression is relaxed locally through fiber buckling, whereas tension propagates along extended rope-like force chains~\cite{Ronceray2016}. These mechanisms explain how active contraction can stiffen a fibrous matrix, but they generally treat the active source or its mechanical coupling to the matrix as persistent. Even for passive inclusions~\cite{van2020,gandikota2020loops}, the stiffness of a disordered network can depend on inclusion spacing as well as volume fraction near a rigidity transition~\cite{MacKintosh2025}. An active spheroid additionally changes the active interfacial area, linker number, and spatial organization of the perturbed network. Neither inclusion volume nor scalar prestress alone is therefore expected to fully determine the composite modulus.

A missing ingredient is the persistence of the spheroid--matrix interface. A solid-like spheroid maintains a comparatively stable boundary topology, allowing cell--matrix linkers to remain coupled to coherent fiber pathways. A fluid-like spheroid instead undergoes more frequent cellular rearrangements, which trigger geometry-driven linker removal and reassignment as the boundary evolves. Because newly assigned linkers can contract and carry force again, this renewal does not simply eliminate prestress. Rather, it relocates the position and orientation of the applied traction and can disrupt the continuity of the pathways that transmit stress during shear. Systems with similar scalar prestress may consequently exhibit different shear moduli. Related vertex-model studies have examined tissue instabilities under cellular activity or applied stresses~\cite{PerezVerdugo2020} and tissue rheology under simple and oscillatory shear~\cite{GrossmanJoanny2025}. Using a vertex model~\cite{Honda1982,Honda2004,Farhadifar_2007,Fletcher2013,Okuda2013,Okuda2015,Bi_2015,Barton2017,Alt2017,sarkar2024graph,Lange2025} to represent a spheroid, our previous work established the coupled three-dimensional vertex model--fiber framework used to study spheroid mechanics~\cite{Zhang2022}, showed that spheroid fluidity regulates long-time matrix displacement and densification~\cite{zhang2025}, and identified state-dependent cellular stress organization within embedded spheroids~\cite{ameen2026}. These findings motivate the complementary question addressed here: how does spheroid fluidity regulate the conversion of active prestress into macroscopic stiffness before extensive matrix remodeling develops?

Here, we address this problem using three-dimensional vertex-model spheroids embedded in a diluted semiflexible fiber network and coupled to nearby fibers by passive or actively contractile linkers whose assignments evolve with the spheroid boundary. We compare solid-like and fluid-like spheroids with matched intact, voided, and passive-linker controls to separate fiber removal, passive interfacial tethering, active prestress, and interfacial renewal. We first develop a theoretical framework that successively incorporates passive inclusion mechanics, prestress-induced stiffening, and the persistence of interfacial load transmission. We then use single-spheroid systems at fixed geometry to examine how linker-generated prestress and spheroid fluidity control the low-strain modulus. Next, we vary the spheroid radius and number to investigate how active interfacial area, inclusion volume, boundary spacing, and the spatial organization of prestressed regions influence the mechanical response. Finally, we extend the analysis to large strain to determine how the dominant load-bearing mechanism evolves with applied deformation. This organization allows us to distinguish the roles of spheroid material state, active loading, interface dynamics, and geometry in the shear mechanics of spheroid--fiber composites.

\section{Methods}

\subsection{Three-dimensional vertex model of the spheroid}

Each spheroid is represented as a three-dimensional confluent cellular aggregate with mechanical energy
\begin{equation}
\begin{split}
E_{\mathrm{VM}}
= {} & K_V\sum_j \left(V_j-V_0\right)^2
+K_A\sum_j \left(A_j-A_0\right)^2 \\
& +\Gamma\sum_{\alpha}\delta_{\alpha B}A_{\alpha},
\end{split}
\label{eq:vertex_model_energy}
\end{equation}
where $V_j$ and $A_j$ are the volume and total surface area of cell $j$, and $V_0$ and $A_0$ are their preferred values. The coefficients $K_V$ and $K_A$ set the corresponding penalties. In the final term, $A_{\alpha}$ is the area of face $\alpha$, $\delta_{\alpha B}=1$ for an exterior face and $0$ otherwise, and $\Gamma$ is the additional boundary-face tension.

A finite spheroid is generated by cutting a spherical cellular aggregate from a bulk confluent tissue. The preferred cell geometry is controlled by the dimensionless target shape index
\begin{equation}
s_0=\frac{A_0}{V_0^{2/3}}.
\label{eq:target_shape_index}
\end{equation}
We use $s_0=5.0$ for solid-like spheroids and $s_0=5.8$ for fluid-like spheroids, as calibrated previously for the same vertex model~\cite{Zhang2022,zhang2025,ameen2026}. Cellular reconnections follow the topology-update rules described in those references.

Lengths are reported in units of $\ell=V_0^{1/3}$ and energies in units of $E_0=K_A V_0^{4/3}$. The overdamped time unit is $t_0=\ell^2/(\mu E_0)$, where $\mu$ is the common mobility of cell vertices and unconstrained fiber nodes. In reduced units, $V_0=1$, $K_A=1$, and $\mu=1$.

\subsection{Disordered fiber network and embedded geometry}

The extracellular matrix is represented by a diluted fiber network constructed from a face-centered-cubic (FCC) lattice. At each FCC site, the six initially straight fiber axes are randomly paired into three coincident, mechanically independent phantom crosslinks. Each crosslink couples at most two fibers, has maximum coordination four, prevents relative sliding, and permits free rotation~\cite{Sharma2016-kf}.

Each fiber bond is retained independently with probability $p$ and removed with probability $1-p$. The realized mean coordination is
\begin{equation}
z=\frac{2N_b}{N_n},
\label{eq:fiber_network_coordination}
\end{equation}
where $N_b$ and $N_n$ are the numbers of retained fiber bonds and phantom nodes, respectively; spheroid--matrix linkers are excluded. Unless otherwise stated, $p=0.91$, giving $z\simeq3.6$.

The fiber-network energy is
\begin{equation}
\begin{split}
E_{\mathrm{FB}}
= {} & \frac{K_S}{2}
\sum_{\langle ij\rangle}
n_{ij}\left(\ell_{ij}-\ell_{\mathrm{f},0}\right)^2 \\
& +\frac{K_B}{2}
\sum_{\langle ijk\rangle_{\mathrm f}}
n_{ij}n_{jk}
\left[1+\cos\left(\theta_{ijk}\right)\right]^2,
\end{split}
\label{eq:fiber_network_energy}
\end{equation}
where $n_{ij}$ is the bond-occupancy variable, $\ell_{ij}$ is the instantaneous bond length, $\ell_{\mathrm{f},0}$ is the undeformed bond length, and $\langle ijk\rangle_{\mathrm f}$ denotes consecutive bond pairs on the same fiber. The angle $\theta_{ijk}$ equals $\pi$ for a straight fiber, so the angular term vanishes there and scales as $K_B(\pi-\theta_{ijk})^4/8$ for small deviations. It is therefore quartic rather than harmonic near the straight state.

The nominal coordinate box has side length $L=32\ell$. Periodic boundary conditions are imposed in the $x$ and $y$ directions. Fiber-node layers at $z=\pm15\ell$ are fixed, giving a separation $H=30\ell$ between the sheared boundaries. Spherical cavities are created by removing fiber nodes inside the prescribed regions and all incident bonds. A single cavity is centered in the box. For multiple-spheroid systems, $N_s$ nonoverlapping cavities are placed with minimum-image center-to-center distances greater than $2R$, and each center is at least $2R$ from the fixed top and bottom boundaries. All matched systems use the same fiber-network realization, cavity positions, and cavity radii.

We compare the active networks with three reference systems. The intact control contains neither cavities nor spheroids. The voided control contains the prescribed cavities but no spheroids or linkers. The passive-linker control contains spheroids and dynamically reassigned linkers, but each linker rest length remains fixed after creation.

For $N_s$ spheroids of prescribed radius $R$, the nominal spheroid volume fraction is
\begin{equation}
\phi=\frac{4\pi N_sR^3}{3V_{\mathrm{box}}},
\qquad
V_{\mathrm{box}}=L^3=(32\ell)^3.
\label{eq:spheroid_volume_fraction_methods}
\end{equation}
For a single spheroid, the surface-to-surface distance from its periodic image is $\xi_p=L-2R$. For multiple spheroids, $\xi_p=\langle d_{\mathrm{nn}}-2R\rangle$, where $d_{\mathrm{nn}}$ is the minimum-image nearest-neighbor center distance. We study single spheroids with $R/\ell=3$, $3.78$, $4.77$, and $6$, and systems containing $N_s=1$, $2$, $4$, or $8$ spheroids with $R/\ell=3$.

\subsection{Dynamic spheroid--fiber linkers}

The spheroid boundary is coupled to the cavity surface by harmonic linker springs. In the implementation, an exterior cell face is represented as a surface polygon. Each eligible cavity-boundary node proposes its nearest valid surface polygon within $2\ell$; proposals are accepted in order of separation, with at most one linker per fiber node, surface polygon, and boundary cell. The target linker number $N_{\mathrm{LS}}^{\mathrm{tar}}$ is chosen proportional to the combined nominal spheroid surface area, holding the target surface density fixed when $R$ or $N_s$ is varied.

Because geometrical constraints can prevent this target from being reached, we distinguish $N_{\mathrm{LS}}^{\mathrm{tar}}$ from the instantaneous linker set $\mathcal L$. The linker energy is
\begin{equation}
E_{\mathrm{LS}}
=\frac{K_{\mathrm{LS}}}{2}
\sum_{i\in\mathcal L}
\left(\ell_{\mathrm{L},i}-\ell_{\mathrm{L},0i}\right)^2,
\label{eq:linker_energy}
\end{equation}
where $\ell_{\mathrm{L},i}$ and $\ell_{\mathrm{L},0i}$ are the instantaneous and rest lengths of linker $i$. A newly assigned linker has $\ell_{\mathrm{L},0i}=\ell_{\mathrm{L},i}$ and is therefore force free. Its rest length then remains fixed in the passive control or follows the same active-contraction rule as the initial linkers.

Linker renewal is geometry mediated: the model includes no force- or extension-dependent rupture criterion. A linker is removed only when its surface polygon is eliminated by a topology update or becomes geometrically invalid. A polygon is invalid if it has fewer than three edges, has area below $0.1\ell^2$, contains an edge shorter than $0.02\ell$ or longer than $1.5\ell$, or if, for any edge, either endpoint angle in the triangle formed by that edge and the polygon center is below $\arccos(0.999)\simeq2.56^{\circ}$. The last condition identifies a nearly degenerate center--edge triangulation. These checks also capture a boundary cell moving into the spheroid interior: disappearance of its surface polygon invalidates the associated linker and releases the fiber node for reassignment. Linker validity is checked after reconnection attempts during overdamped evolution and every 100 FIRE iterations. The pairing routine then attempts to restore $N_{\mathrm{LS}}^{\mathrm{tar}}$ from the remaining eligible nodes and valid polygons; if too few pairs are available, the instantaneous linker number remains below the target.

\begin{figure*}[t]
\centering
\includegraphics[width=0.85\linewidth]{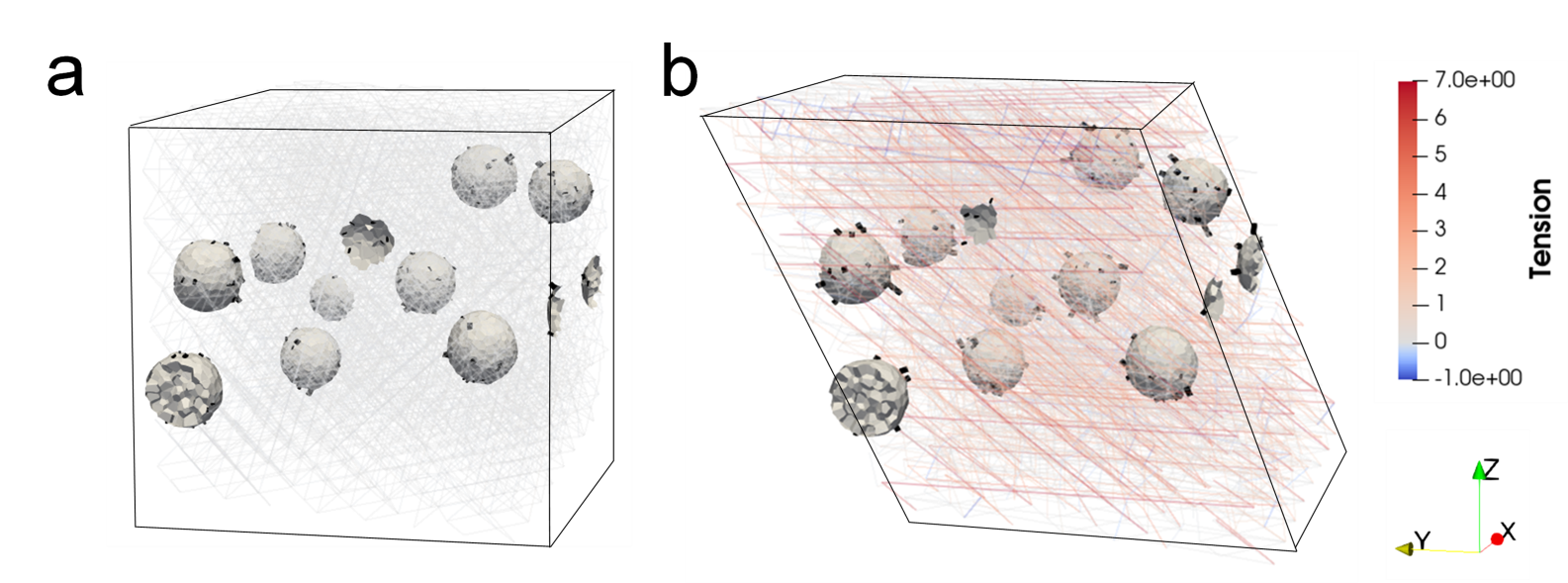}
\caption{\textbf{Deformation of a fiber network containing eight embedded spheroids under shear.} Simulation snapshots at (a) $0\%$ and (b) $50\%$ applied shear strain. The spheroid is shown in white, spheroid--matrix linkers in black, and fiber color denotes the stretching force.}
\label{fig:deformation_single_network}
\end{figure*}

\subsection{Preparation, active contraction, and shear}

The mechanical energy used to calculate forces is
\begin{equation}
E_{\mathrm{ES}}
=E_{\mathrm{VM}}+E_{\mathrm{FB}}+E_{\mathrm{LS}}.
\label{eq:total_mechanical_energy}
\end{equation}
Before active contraction and shear, the coupled system is evolved for $600t_0$ by deterministic overdamped dynamics with time step $\Delta t=0.005t_0$ and no Brownian forcing. The equations of motion are
\begin{equation}
\dot{\mathbf r}_I
=-\mu\frac{\partial E_{\mathrm{ES}}}{\partial\mathbf r_I},
\qquad
\dot{\mathbf R}_i
=-\mu\frac{\partial E_{\mathrm{ES}}}{\partial\mathbf R_i},
\label{eq:overdamped_dynamics}
\end{equation}
for cell vertex $I$ and unconstrained fiber node $i$, respectively. Cellular reconnections, linker-validity checks, and reassignment remain enabled during this pretreatment, but linker rest lengths are fixed, so active contraction does not yet occur.

The pretreated system is next relaxed at zero applied strain using the fast inertial relaxation engine (FIRE)~\cite{GUENOLE2020109584}, during which active contraction begins. At FIRE iteration $m$, each eligible active-linker rest length is updated as
\begin{equation}
\ell_{\mathrm{L},0i}^{(m+1)}
=\ell_{\mathrm{aim}}
+q\left[
\ell_{\mathrm{L},0i}^{(m)}-\ell_{\mathrm{aim}}
\right],
\label{eq:active_linker_update}
\end{equation}
with $q=0.99$ in the standard protocol. The update is applied only while $\ell_{\mathrm{L},0i}^{(m)}>\ell_{\mathrm{aim}}+10^{-3}\ell$ and $\left|\ell_{\mathrm{L},i}-\ell_{\mathrm{L},0i}^{(m)}\right|\leq0.2\ell$. Exceeding the latter threshold pauses shortening but does not remove the linker. Unless otherwise stated, $\ell_{\mathrm{aim}}=0.20\ell$; the prestress scan varies this target to generate different relaxed prestresses. Because Eq.~\eqref{eq:active_linker_update} is indexed by FIRE iterations, it specifies an algorithmic contraction--relaxation protocol rather than an independent physical contraction time.

In the contraction-protocol controls, the final target is unchanged while the rest-length decrement per FIRE iteration is reduced by factors of two and four. These protocols correspond to $q=0.995$ and $q=0.9975$ and are denoted $2\times$ slow and $4\times$ slow, respectively.

After the relaxed $\gamma=0$ state is recorded, $yz$ shear is applied in increments of $\Delta\gamma=5\times10^{-3}$. At each increment, the fixed fiber nodes at $z=\pm H/2=\pm15\ell$ are displaced in the $y$ direction by $\pm\Delta\gamma H/2$. The box and internal degrees of freedom are not transformed affinely; instead, unconstrained fiber nodes and cell vertices are relaxed with FIRE after each boundary displacement. Linker-validity checks, reassignment, and active rest-length updates remain enabled, and cellular reconnections are attempted every 1000 FIRE iterations while active shortening is ongoing. Only FIRE relaxation is used during shear. Loading continues to $\gamma=0.50$.

For the full spheroid model, convergence is assessed every 1000 FIRE iterations from the change in the fiber stretching-plus-angular energy, $E_{\mathrm{FB}}$, and is satisfied when
\begin{equation}
\left|\Delta E_{\mathrm{FB}}\right|<10^{-5}E_0
\quad\text{or}\quad
\frac{\left|\Delta E_{\mathrm{FB}}\right|}
{\left|E_{\mathrm{FB}}^{\mathrm{prev}}\right|}<10^{-3}.
\label{eq:fire_convergence}
\end{equation}
Relaxation terminates when this energy criterion is satisfied and no active-linker rest length is shortened in the current update. No separate residual-force or total-energy tolerance is imposed. The maximum FIRE time step is $0.015t_0$.

Unless otherwise stated, the reduced parameters are
\begin{equation}
\begin{gathered}
K_V=10.0,\quad K_A=1.0,\quad \Gamma=1.0,\\
K_S=10.0,\quad K_B=0.001,\quad K_{\mathrm{LS}}=1.0,\\
\ell_{\mathrm{f},0}=2\sqrt{2}\,\ell,\quad \mu=1.0.
\end{gathered}
\label{eq:simulation_parameters}
\end{equation}
The cellular reconnection threshold is $0.02\ell$.

\subsection{Mechanical and structural observables}

Post-processing quantifies the external fiber network rather than the total spheroid--linker composite. Fiber energies and virial stresses are normalized by the fixed factor
\begin{equation}
V_{\mathrm{norm}}=(30\ell)^3=27000\ell^3.
\label{eq:mechanical_normalization_volume}
\end{equation}
This post-processing normalization differs from the nominal volume $V_{\mathrm{box}}=(32\ell)^3$ used to define $\phi$ in Eq.~\eqref{eq:spheroid_volume_fraction_methods}.

The quantity denoted by $G$ is the fiber-energy-derived apparent shear modulus
\begin{equation}
G(\gamma)
=\frac{1}{V_{\mathrm{norm}}\gamma}
\frac{dE_{\mathrm{FB}}}{d\gamma},
\qquad \gamma>0.
\label{eq:apparent_shear_modulus}
\end{equation}
For each realization, the derivative is calculated with \texttt{numpy.gradient} on the discrete relaxed energy--strain series, using centered differences at interior points and one-sided differences at the endpoints; the resulting moduli are then ensemble averaged. Because only $E_{\mathrm{FB}}$ is included and the interface can continue to evolve during loading, $G$ is a matrix-only, finite-amplitude, protocol-dependent apparent modulus rather than an equilibrium differential modulus.

The scalar prestress is evaluated from the recorded $\gamma=0$ configuration after active contraction and FIRE relaxation and before the first shear increment. It is the trace of the stretching virial stress of the external fiber bonds and is implemented as
\begin{equation}
\sigma
=\operatorname{tr}\left(\boldsymbol{\sigma}^{\mathrm s}\right)
=-\frac{1}{V_{\mathrm{norm}}}
\sum_{\langle ij\rangle}T_{ij}\ell_{ij},
\label{eq:fiber_prestress_definition}
\end{equation}
where $T_{ij}=K_S(\ell_{ij}-\ell_{\mathrm{f},0})$ is the signed axial force and $\ell_{ij}$ is evaluated from the minimum-image bond vector. Each external fiber bond is counted once. A stretched bond has $T_{ij}>0$ and contributes negatively, so tensile contractile states have $\sigma<0$. The trace is not divided by the spatial dimension; angular, linker, cellular, and fixed-topology-inclusion contributions are excluded.

The simulation records the sum of squared linker-force components on the linked fiber nodes, $S_{\mathrm L}=\sum_a|\mathbf F_a^{\mathrm L}|^2$. The reported quantity is the reference-normalized root-mean-square Cartesian linker-force component
\begin{equation}
F_{\mathrm L}
=\left[
\frac{S_{\mathrm L}}{3N_{\mathrm{LS}}^{\mathrm{ref}}}
\right]^{1/2},
\label{eq:linker_force_observable}
\end{equation}
where $N_{\mathrm{LS}}^{\mathrm{ref}}=80$ is fixed rather than replaced by the instantaneous linker number when the linker set changes.

Linker-update activity is obtained from the counter for new node--polygon assignments, which is reset at the beginning of each FIRE relaxation and after every log record. If $n_{kq}^{\mathrm{new}}$ is the number of new assignments in logging interval $q$ of the stage at strain $\gamma_k$, the reported activity is
\begin{equation}
A_{\mathrm L}(\gamma_k)
=\frac{1}{M_k}\sum_{q=1}^{M_k}n_{kq}^{\mathrm{new}},
\label{eq:linker_update_activity}
\end{equation}
where $M_k$ is the number of logged intervals in that FIRE stage. The initial linker set is excluded, while every repeated reassignment contributes a new event. Thus, $A_{\mathrm L}$ is an interval-averaged, protocol-dependent activity, not cumulative turnover, force-induced rupture, a physical rate, or a survival fraction. Solid-like and fluid-like systems are processed identically.

For the single-spheroid radial profiles, each bond center is assigned its minimum-image distance from the spheroid center; the prescribed radius is subtracted, and $T_{ij}$ is averaged in common surface-distance bins. The global fiber-tension observable is the signed sum $\mathcal T_{\mathrm{FB}}=\sum_{\langle ij\rangle}n_{ij}T_{ij}$ over occupied external-fiber bonds, evaluated for each realization before ensemble averaging. A consecutive bond pair is classified as buckled when $\theta_{ijk}<150^{\circ}$, in which case both adjacent bonds are marked. For a single spheroid, $R_{\mathrm b}$ is the 75th percentile of the marked-bond-center distances from the spheroid center, i.e., the radius containing $75\%$ of the buckled bonds.

The global nonaffine displacement of the external network is
\begin{equation}
U_{\mathrm{NA}}(\gamma)
=\frac{1}{N_{\mathrm f}}
\sum_{i\in\mathrm{free}}
\left|
\Delta\mathbf R_i^{\mathrm{MI}}(\gamma)
-\gamma z_i^{\mathrm{ref}}\mathbf e_y
\right|,
\label{eq:global_nonaffinity}
\end{equation}
where $N_{\mathrm f}$ is the number of unconstrained fiber nodes, the first analyzed frame supplies $z_i^{\mathrm{ref}}$, and $\Delta\mathbf R_i^{\mathrm{MI}}$ is the displacement from that frame with minimum-image corrections in $x$ and $y$. The subtracted affine field is the absolute imposed field $\gamma z_i^{\mathrm{ref}}\mathbf e_y$. Thus, $U_{\mathrm{NA}}$ is a mean displacement magnitude, not a squared or strain-normalized measure. The fiber-translation maps show total, rather than affine-subtracted, displacements relative to the first analyzed frame; node displacements are binned by their reference positions in the $yz$ plane and averaged within bins and across realizations. The separately reported stretching and angular energies are the first and second terms of Eq.~\eqref{eq:fiber_network_energy}.

For the cellular analysis, the maximum shear stress of cell $j$ is calculated from the largest and smallest principal values of its stress tensor as
\begin{equation}
\tau_{\max,j}=\frac{\sigma_{\max,j}-\sigma_{\min,j}}{2}.
\label{eq:cell_maximum_shear}
\end{equation}
The cell shape tensor is constructed from the positions of the cell vertices and polygon centers relative to the cell center. If its ordered eigenvalues are $\lambda_1\leq\lambda_2\leq\lambda_3$, the shape anisotropy is
\begin{equation}
\Delta
=\frac{
\left[\lambda_3-\left(\lambda_1+\lambda_2\right)/2\right]^2
+3\left(\lambda_2-\lambda_1\right)^2/4
}{
\left(\lambda_1+\lambda_2+\lambda_3\right)^2
}.
\label{eq:cell_shape_anisotropy}
\end{equation}
Cell layers are assigned by iteratively removing the current exterior-cell shell. The implementation indexes the outermost shell as 0; for presentation, this index is shifted by one so that layer 1 is the outermost boundary-cell layer and the layer number increases inward. Cellular stress and shape tensors follow the procedures used for the same vertex model in Refs.~\cite{Zhang2022,ameen2026}.

Unless otherwise stated, ensemble curves report the mean and standard error over independent realizations; the main ensembles contain $100$ realizations where indicated. Layer-resolved cellular profiles instead report the mean and sample standard deviation. The low-strain modulus is evaluated at $\gamma_m=0.01$. The solid-like prestress data are fitted to
\begin{equation}
G(\gamma_m)
=G_{\mathrm{pass}}(\gamma_m)
+C_{\mathrm{pre}}|\sigma|^{\alpha},
\label{eq:prestress_fit_methods}
\end{equation}
where $C_{\mathrm{pre}}$ is the prestress-stiffening prefactor and $\alpha$ is the fitted exponent. In the power-law fit, $G_{\mathrm{pass}}$ is a fitted offset; in the turnover analysis, it is the modulus of the matched passive-linker system for each spheroid state. Further fitting choices are specified in the corresponding figure captions.

\subsection{Fixed-topology spring-network control}

To separate the direct effect of inclusion rigidity from cellular rearrangement and linker reassignment, the vertex-model spheroid is replaced by a spherical, nonphantom FCC spring network with fixed connectivity. The inclusion has stretching interactions only,
\begin{equation}
E_{\mathrm I}
=\frac{K_{\mathrm I}}{2}
\sum_{\langle ij\rangle\in\mathcal B_{\mathrm I}}
\left(\ell_{ij}-\ell_{\mathrm I,0}\right)^2,
\label{eq:toy_inclusion_energy}
\end{equation}
where $\mathcal B_{\mathrm I}$ is the set of inclusion bonds. The inclusion contains no angular elasticity or topology changes; $K_{\mathrm I}=10$ and $\ell_{\mathrm I,0}=2\sqrt{2}\,\ell$. Its realized connectivity is
\begin{equation}
z_{\mathrm I}=\frac{2N_{b,\mathrm I}}{N_{n,\mathrm I}},
\label{eq:toy_inclusion_connectivity}
\end{equation}
where $N_{b,\mathrm I}$ and $N_{n,\mathrm I}$ are the numbers of inclusion bonds and nodes, respectively; spheroid--matrix linkers are excluded. We compare a floppy inclusion with $z_{\mathrm I}=4.0$ and a rigid inclusion with $z_{\mathrm I}\simeq8$.

The paired floppy and rigid systems use the same external voided network, inclusion-node positions, and 80 permanent linker endpoint pairs. Each linker is initially force free. The noncontractile control retains its initial rest length. In the active control, Eq.~\eqref{eq:active_linker_update} is used with $\ell_{\mathrm{aim}}=\ell_{\mathrm{L},0}^{\mathrm{tar}}=0.20\ell$, $q=0.999$, and a stopping tolerance of $0.01\ell$. The assembled external network, inclusion, and linkers are relaxed together at zero strain; the inclusion is not inserted into a separately prerelaxed voided network.

The toy model uses a force-based FIRE criterion: relaxation ends after $F_{\mathrm{toy}}^2<10^{-10}$ has been recorded on more than $1000$ iterations and active shortening has ceased, where $F_{\mathrm{toy}}^2$ is the sum of squared total-force components over unconstrained nodes. During shear, the inclusion connectivity, linker endpoint identities, and linker rest lengths remain fixed, while all unconstrained inclusion and external-network nodes relax with FIRE; linker removal and reassignment are disabled. Shear is applied in increments of $\Delta\gamma=0.005$ to $\gamma=0.10$. The apparent modulus and prestress include only the external fiber bonds [Eqs.~\eqref{eq:apparent_shear_modulus} and \eqref{eq:fiber_prestress_definition}], excluding inclusion and linker energies and stresses. Paired realizations are summarized in Fig.~\ref{fig:supp_toy_model_simulations}.

Further details of the three-dimensional vertex construction and cellular reconnection algorithms are provided in Refs.~\cite{Zhang2022,zhang2025,ameen2026}.

\section{Results}

\subsection{Theoretical framework for spheroid--fiber composite mechanics}
\label{subsec:toy_model}

\begin{figure}[t]
\centering
\includegraphics[width=\columnwidth]{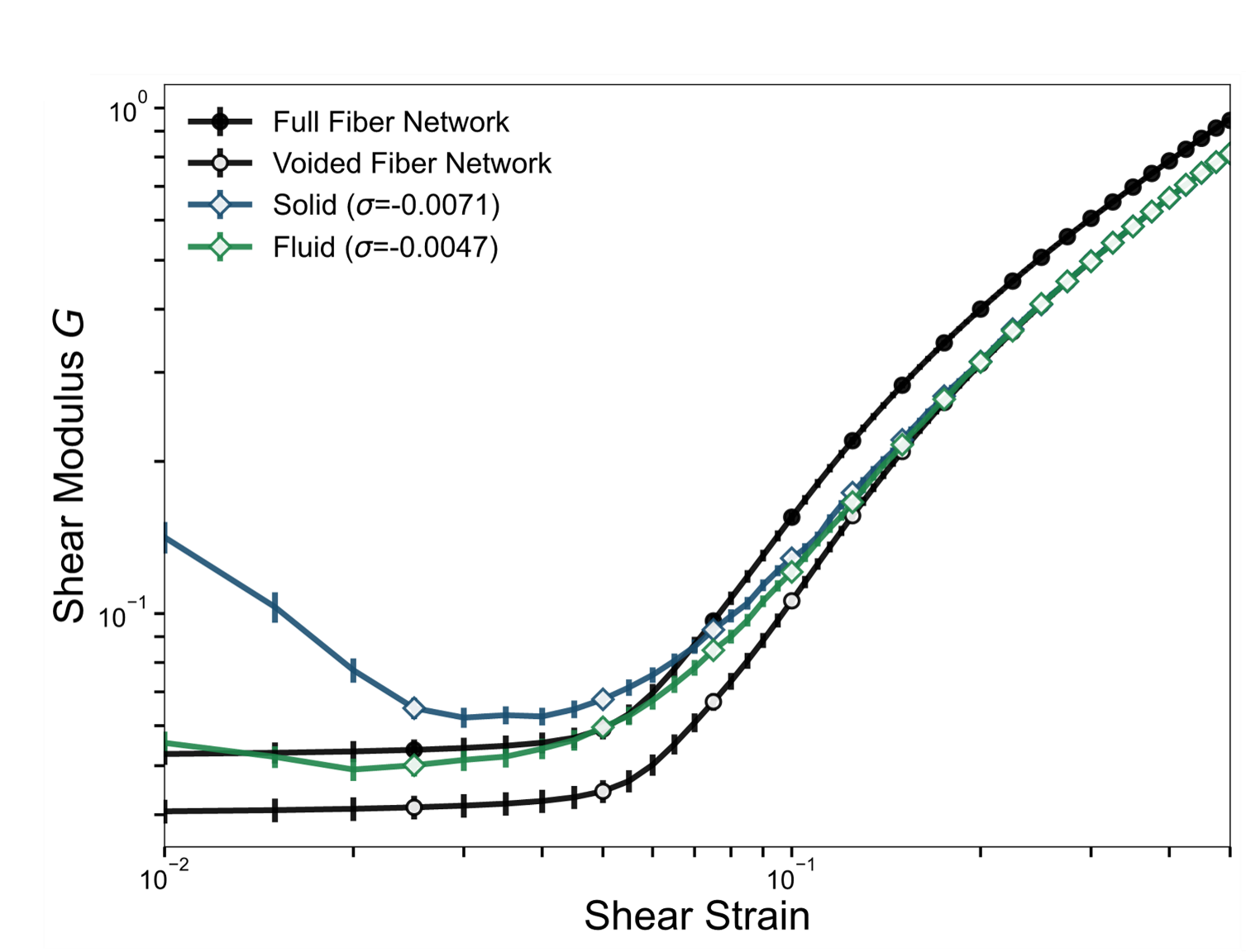}
\caption{\textbf{Overview of the strain-dependent mechanical response of spheroid--fiber composites.} Apparent shear modulus $G$ as a function of applied shear strain $\gamma$ for an intact fiber network, a matched eight-void fiber network, and networks containing eight solid-like or fluid-like spheroids coupled to the surrounding fibers by active contractile linkers. Each spheroid and corresponding cavity has radius $R=3$. The intact network contains no cavities, whereas the eight-void reference contains the same cavity geometry as the spheroid-containing systems but no spheroids or spheroid--matrix linkers, thereby isolating the response of the residual fiber backbone. The values of $\sigma$ in the legend denote the fiber prestress at $\gamma=0$, measured after active contraction and mechanical relaxation but before shear is applied; negative values correspond to contractile states. Each curve represents an average over $100$ independent simulations.}
\label{fig:large_strain_converge}
\end{figure}

The mechanical response of the spheroid--fiber composite arises from coupled material, interfacial, and geometric factors. Spheroid rigidity controls the ability of an inclusion to support shear, while the spheroid radius and number jointly determine the occupied volume, total active interfacial area, number of linkers, and spacing and overlap between mechanically perturbed regions. Active linker contraction generates prestress, whereas linker turnover controls the persistence and spatial organization of interfacial load transmission. These contributions act through a nonlinear fiber network whose response changes with applied strain.

Figure~\ref{fig:large_strain_converge} presents a representative response for $N_s=8$ spheroids of radius $R=3$. At this fixed geometry, solid-like and fluid-like spheroids exhibit distinct low-strain moduli and different pre-shear prestresses, whereas both systems approach the matched voided-network response at large strain. This behavior illustrates the coupled nature of the composite mechanics but does not isolate the individual contributions. We therefore construct a hierarchy of limiting cases that successively considers passive spheroid reinforcement, active prestress, and linker turnover. Unless stated otherwise, all moduli below refer to the low-strain apparent modulus evaluated at the same measurement strain $\gamma_m$.

We first consider passive and permanent linkers. Their stiffness and density are fixed, and no active prestress is generated. Let $G_m$ be the modulus of the inclusion-free fiber matrix, $G_c$ the intrinsic shear modulus of the spheroid, and $G_{\mathrm{void}}$ the modulus of the matched voided network. The relative spheroid rigidity is
\begin{equation}
\eta = \frac{G_c}{G_m}.
\label{eq:relative_spheroid_rigidity}
\end{equation}
A fluid-like spheroid has $\eta$ close to zero, whereas a rigid spheroid has $\eta$ much larger than one. In the dilute limit, the passive modulus may be written as
\begin{equation}
G_{\mathrm{pass}} \simeq G_{\mathrm{void}} [1 + \phi B(\eta)].
\label{eq:passive_composite_modulus}
\end{equation}
Here, $\phi$ is the total spheroid volume fraction, which depends on both spheroid radius and number. The function $B(\eta)$ describes passive reinforcement. Its fluid and rigid limits are denoted by $B_0$ and $B_\infty$. A simple interpolation is
\begin{equation}
B(\eta) = B_0 + (B_\infty - B_0) \frac{\eta}{1 + \eta}.
\label{eq:passive_rigidity_interpolation}
\end{equation}
The value $B_0$ need not be zero because a fluid-like spheroid remains mechanically coupled to the matrix. If $B_\infty>B_0$, solid-like spheroids provide stronger passive reinforcement. The leading difference is proportional to $\phi$ in the dilute limit.

We next allow the permanent linkers to contract and generate prestress. Let $\sigma$ be the scalar fiber prestress measured before shear. Contractile states have negative $\sigma$ under our sign convention, so $|\sigma|$ denotes the prestress magnitude. At fixed geometry, a minimal low-strain relation is
\begin{equation}
G - G_{\mathrm{pass}} \simeq C_{\mathrm{pre}}(\eta) |\sigma|^\alpha.
\label{eq:permanent_linker_stiffening}
\end{equation}
Here, $C_{\mathrm{pre}}(\eta)$ measures how efficiently prestress is converted into additional shear stiffness, and $\alpha$ is an effective stress-stiffening exponent. For a given spheroid type and geometry, $C_{\mathrm{pre}}(\eta)$ denotes the corresponding permanent-interface conversion coefficient. It depends on spheroid rigidity and can also depend on active-boundary geometry, spacing, and overlap between mechanically perturbed regions. Because $|\sigma|$ is the system-averaged matrix prestress, it already contains the effects of spheroid number, active area, and linker activity on the magnitude of active loading. These quantities are therefore not introduced as additional simple prefactors. Their spatial organization can nevertheless affect $C_{\mathrm{pre}}$.

The nonlinear active-inclusion theory of Ronceray and coworkers provides a possible interpretation of this prestress dependence.\cite{Ronceray2019} For a spherical active source of radius $R$, the theoretical nonlinear-region radius $R_*$ satisfies
\begin{equation}
R_* = R [ \frac{\sigma_a}{d (\sigma_{\mathrm{ext}} + \sigma_b)} ]^{1/(d-1)}.
\label{eq:nonlinear_region_radius}
\end{equation}
Here, $\sigma_a$ is the local active source stress, $\sigma_b$ is the buckling threshold, $\sigma_{\mathrm{ext}}$ is an independently imposed isotropic stress, and $d$ is the spatial dimension. These stresses are different from the measured global prestress $\sigma$. The theoretical radius $R_*$ is also different from the simulation-defined buckling radius $R_b$ introduced below.

If the nonlinear regions are dilute and do not overlap, a simple volume estimate gives
\begin{equation}
G - G_{\mathrm{pass}} \simeq \phi \Delta G_{\mathrm{nl}} [ (\frac{R_*}{R})^3 - 1 ].
\label{eq:nonlinear_region_modulus}
\end{equation}
Here, $\Delta G_{\mathrm{nl}}$ is the excess local modulus of the nonlinear region. This estimate applies only when $R_*>R$ and $R_*\ll L$. If $\sigma_a$ is proportional to $|\sigma|$ at fixed geometry, the three-dimensional volume estimate suggests an effective exponent close to $3/2$. This exponent is not a direct prediction of the active-inclusion theory because the volume-to-modulus relation is an additional assumption. The estimate also fails when nonlinear regions surrounding neighboring spheroids overlap.

We finally include geometry-driven linker turnover. The prestress magnitude $|\sigma|$ is measured at $\gamma=0$, after zero-strain contraction and relaxation but before the first shear increment. It therefore includes the effect of any linker turnover that occurs during the pre-shear preparation, but not turnover during the subsequent shear measurement.

During shear, removal of a loaded linker temporarily releases local stress. Its replacement is initially force free, but continued active contraction can generate new stress, usually at a different position and orientation. To obtain a minimal closure, we neglect a separate prestress-retention factor over the low-strain measurement interval. This approximation assumes that the temporary reduction in scalar prestress caused by linker removal is substantially rebuilt by subsequent contraction. We therefore focus on the remaining effect of turnover: the relocation of traction and the resulting disruption of coherent interfacial load paths. We introduce an effective path-transmission factor $P_L$ to describe this effect. The factor includes contributions from both surviving linkers and renewed linkers that reload during shear; it is not identified with the survival fraction alone.

We quantify turnover using the normalized cumulative renewal count,
\begin{equation}
\chi_L(\gamma_m) = \frac{N_{\mathrm{new}}(0,\gamma_m)}{N_{\mathrm{LS}}(0)}.
\label{eq:normalized_linker_turnover}
\end{equation}
Here, $N_{\mathrm{new}}(0,\gamma_m)$ is the number of newly assigned linkers accumulated from $\gamma=0$ to the measurement strain. For independent and spatially uniform renewal events, $\exp(-\chi_L)$ would approximate the survival fraction of the initial linkers. Survival is not identical to path transmission, however, because newly assigned linkers can contract and establish new load-bearing paths. We therefore introduce a dimensionless, state-dependent path-disruption coefficient $\lambda(\eta)$ and use the lowest-order closure
\begin{equation}
P_L(\gamma_m,\eta) \simeq \exp[-\lambda(\eta) \chi_L(\gamma_m)].
\label{eq:path_transmission_factor}
\end{equation}
The coefficient $\lambda(\eta)$ measures the average loss of coherent transmission produced by a normalized linker-renewal event after allowing renewed linkers to contract and reload. The same number of renewal events need not have the same mechanical consequence for a stable solid-like boundary and a rearranging fluid-like boundary. We denote the corresponding values by $\lambda_{\mathrm{s}}$ and $\lambda_{\mathrm{f}}$, without assuming a particular interpolation between them. The limit $\lambda=0$ corresponds to complete recovery of path transmission, values between zero and one describe partial recovery, $\lambda=1$ gives the survival-like estimate, and values larger than one represent stronger collective disruption of load-bearing paths. Within this loss-only closure, $\lambda$ is nonnegative. An unconstrained fit giving $\lambda<0$ should therefore not be interpreted as negative disruption; it instead indicates that renewal is mechanically neutral or adaptive, or that the renewal count is not a sufficient proxy for path transmission. The coefficient is not a physical turnover time or rate. It is treated as constant only within a given spheroid state, matched geometry, and measurement protocol, and no common value is assumed for solid-like and fluid-like spheroids.

The resulting low-strain prediction is
\begin{equation}
G(\gamma_m) - G_{\mathrm{pass}}(\gamma_m) \simeq C_{\mathrm{pre}}(\eta) \exp[-\lambda(\eta) \chi_L(\gamma_m)] |\sigma|^\alpha.
\label{eq:turnover_modulus}
\end{equation}
In this equation, $G_{\mathrm{pass}}(\gamma_m)$ is the modulus of the matched passive system with the same spheroid type, geometry, and linker-update rule. Permanent linkers have $\chi_L=0$ and therefore $P_L=1$. For solid-like spheroids, renewed linkers can reload within a comparatively stable boundary structure and along mechanically compatible paths. The effective value $\lambda_{\mathrm{s}}$ may therefore be small even when a finite number of linker updates occurs, recovering the relation $G-G_{\mathrm{pass}}\sim|\sigma|^\alpha$. For fluid-like spheroids, boundary rearrangements can make each renewal event more disruptive in addition to increasing the number of events. Consequently, $\lambda_{\mathrm{f}}$ can differ from, and may exceed, $\lambda_{\mathrm{s}}$. A fluid-like system can then retain substantial prestress while converting it less efficiently into shear stiffness. If the combined quantity $\lambda_{\mathrm{f}}\chi_L$ grows sufficiently rapidly with $|\sigma|$, the exponential loss of path transmission can outweigh the increase in $|\sigma|^\alpha$ and produce a nonmonotonic modulus--prestress relation.

The final equation is a phenomenological approximation, not a statement that prestress, spheroid rigidity, geometry, and interfacial transmission are independent. In particular, it neglects an independent change in prestress magnitude during the low-strain measurement and assigns the effect of turnover only to path transmission. If stress rebuilding is incomplete on this interval, an additional prestress-retention variable would be required. Moreover, $\chi_L$ measures how often linkers are reassigned, whereas $\lambda(\eta)$ represents the mechanical consequence of those events. In general, $C_{\mathrm{pre}}$, $P_L$, and $\lambda$ can depend on spheroid state, geometry, activity, and measurement strain. The state-dependent exponential closure is therefore a phenomenological hypothesis rather than a demonstrated universal collapse. The present data can motivate separate effective values $\lambda_{\mathrm{s}}$ and $\lambda_{\mathrm{f}}$, but do not determine a continuous dependence on $\eta$. Together, these limiting cases organize the low-strain response in terms of passive inclusion mechanics, active prestress, and persistent interfacial transmission. The large-strain convergence in Fig.~\ref{fig:large_strain_converge} is controlled mainly by deformation of the residual fiber backbone and is discussed separately below.

\subsection{Prestress-mediated stiffening by active linkers in single-spheroid networks}
\label{subsec:single_spheroid_prestress}

\begin{table}[t]
\centering
\scriptsize
\setlength{\tabcolsep}{3pt}
\caption{\textbf{Prestress values for spheroid simulations.}
The scalar prestress $\sigma$ is measured in the recorded $\gamma=0$ configuration following zero-strain FIRE contraction and relaxation, before the first boundary-shear increment, as the trace of the stretching virial stress tensor of the fiber network. In Fig.~\ref{fig:single_spheroid_response}, the spheroid radius is fixed at $R=4.77$ and $\sigma$ is tuned by changing the target linker rest length or by using passive linkers. In Fig.~\ref{fig:single_radius_increase}, the active-linker parameters are fixed while the single-spheroid radius $R$ is varied. In Fig.~\ref{fig:multiple_spheroid_mechanics}, the spheroid radius is fixed while the number of spheroids is varied.}
\label{tab:single_spheroid_prestress}
\begin{tabular}{llc}
\hline
Spheroid type & Control parameter & Prestress $\sigma$ \\
\hline
\multicolumn{3}{l}{\textit{Prestress scan, Fig.~\ref{fig:single_spheroid_response}}} \\
Solid-like & Active, high $|\sigma|$ & $-3.12\times10^{-3}$ \\
Solid-like & Active, intermediate $|\sigma|$ & $-2.74\times10^{-3}$ \\
Solid-like & Active, low $|\sigma|$ & $-2.11\times10^{-3}$ \\
Solid-like & Passive & $-3.30\times10^{-5}$ \\
Fluid-like & Active, high $|\sigma|$ & $-2.36\times10^{-3}$ \\
Fluid-like & Active, intermediate $|\sigma|$ & $-2.29\times10^{-3}$ \\
Fluid-like & Active, low $|\sigma|$ & $-1.34\times10^{-3}$ \\
Fluid-like & Passive & $-3.33\times10^{-5}$ \\
\hline
\multicolumn{3}{l}{\textit{Single-spheroid radius scan, Fig.~\ref{fig:single_radius_increase}}} \\
Solid-like & $R=3$ & $-8.34\times10^{-4}$ \\
Solid-like & $R=3.78$ & $-1.67\times10^{-3}$ \\
Solid-like & $R=4.77$ & $-2.11\times10^{-3}$ \\
Solid-like & $R=6$ & $-1.13\times10^{-2}$ \\
Fluid-like & $R=3$ & $-4.94\times10^{-4}$ \\
Fluid-like & $R=3.78$ & $-1.29\times10^{-3}$ \\
Fluid-like & $R=4.77$ & $-1.34\times10^{-3}$ \\
Fluid-like & $R=6$ & $-1.24\times10^{-2}$ \\
\hline
\multicolumn{3}{l}{\textit{Multiple-spheroid scan, Fig.~\ref{fig:multiple_spheroid_mechanics}}} \\
Solid-like & $1$ spheroid & $-1.01\times10^{-3}$ \\
Solid-like & $2$ spheroids & $-2.01\times10^{-3}$ \\
Solid-like & $4$ spheroids & $-3.77\times10^{-3}$ \\
Solid-like & $8$ spheroids & $-7.14\times10^{-3}$ \\
Fluid-like & $1$ spheroid & $-8.78\times10^{-4}$ \\
Fluid-like & $2$ spheroids & $-1.47\times10^{-3}$ \\
Fluid-like & $4$ spheroids & $-2.39\times10^{-3}$ \\
Fluid-like & $8$ spheroids & $-4.66\times10^{-3}$ \\
\hline
\end{tabular}
\end{table}

\begin{figure*}[t]
\centering
\includegraphics[width=0.85\linewidth]{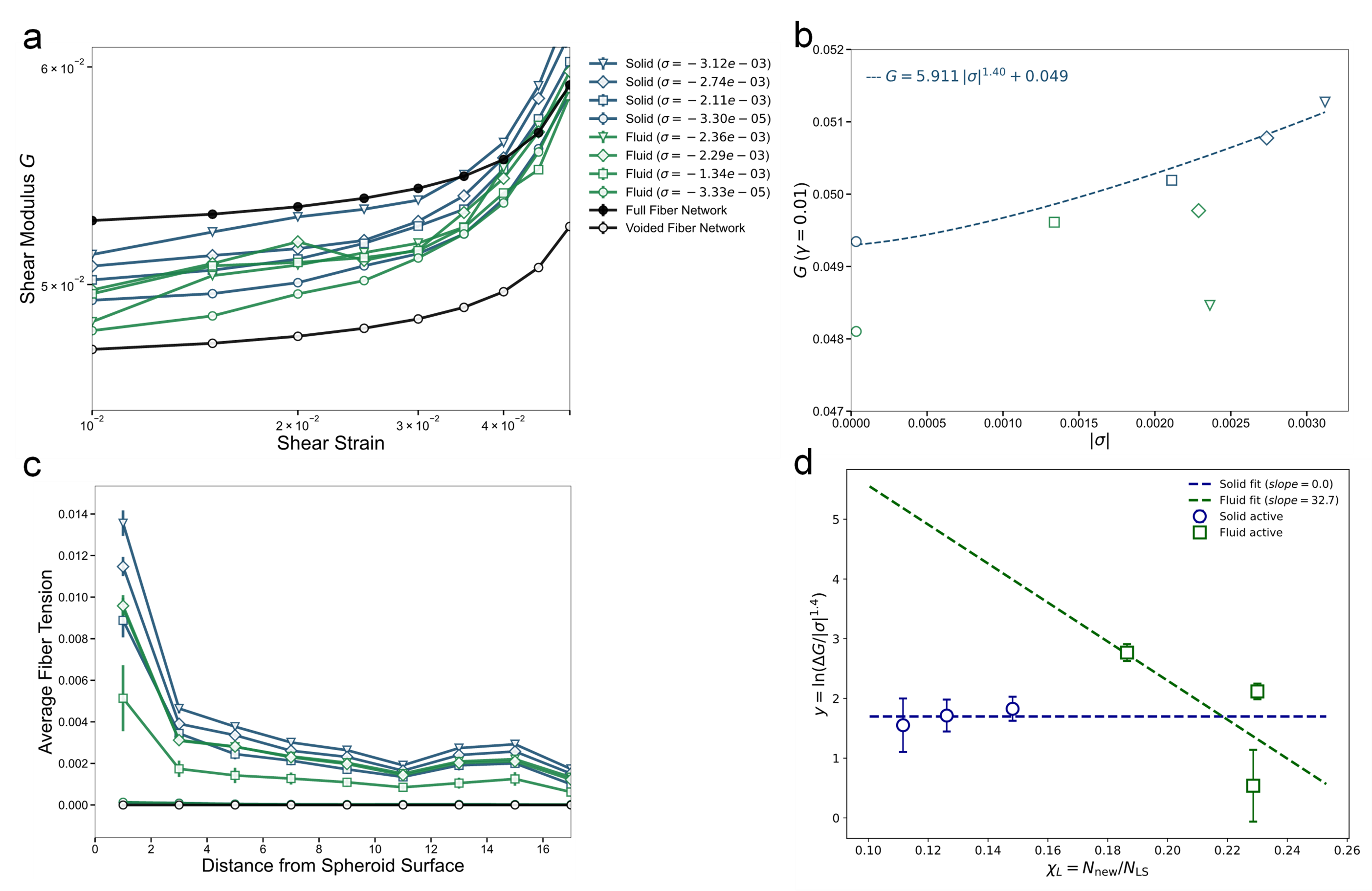}
\caption{\textbf{Prestress-mediated stiffening and state-dependent interfacial transmission in single-spheroid networks.}
(a) Apparent shear modulus $G$ as a function of shear strain $\gamma$ in the small-strain regime for an intact fiber network, a matched voided network, and networks containing a solid-like or fluid-like spheroid with different levels of linker-generated prestress. Open circular markers denote passive-linker systems; the remaining spheroid-containing cases have active linkers, with prestress tuned by varying the target linker rest length.
(b) Apparent shear modulus at $\gamma_m=0.01$ plotted against the prestress magnitude $|\sigma|$. The dashed curve is a fit to the solid-like active-linker data using Eq.~\eqref{eq:permanent_linker_stiffening}, giving $\alpha\simeq1.4$. Here, $G_{\mathrm{pass}}$ is the fitted offset and is close to the modulus of the matched passive solid-like system.
(c) Average pre-shear fiber tension as a function of vertical distance from the spheroid surface.
(d) State-dependent test of Eq.~\eqref{eq:turnover_modulus}. For each spheroid type, $\Delta G=G(\gamma_m)-G_{\mathrm{pass}}(\gamma_m)$ is calculated using its matched passive system, and $y=\ln(\Delta G/|\sigma|^{1.4})$ is plotted against the normalized cumulative linker turnover $\chi_L$ defined by Eq.~\eqref{eq:normalized_linker_turnover}. The fits use $y=a_{\mathrm{type}}-\lambda_{\mathrm{type}}\chi_L$ with the constraint $\lambda_{\mathrm{type}}\geq0$. The constrained solid-like fit reaches the boundary value $\lambda_{\mathrm{s}}=0$, whereas the fluid-like fit gives $\lambda_{\mathrm{f}}=32.7$. All data points represent averages over $100$ simulations.}
\label{fig:single_spheroid_response}
\end{figure*}

We first test the hierarchy proposed in Sec.~\ref{subsec:toy_model} at fixed inclusion geometry. A single spheroid of radius $R=4.77$ is embedded in the fiber network, and the magnitude of active prestress is varied through the target linker rest length; matched passive-linker systems provide the low-prestress baselines listed in Table~\ref{tab:single_spheroid_prestress}. Figure~\ref{fig:single_spheroid_response}(a) shows the resulting small-strain response. The voided network is the most compliant because the empty cavity permits large translations and nonaffine relaxation of the surrounding fibers. Passive spheroid--matrix linkers partially suppress this void-like mode and modestly increase the modulus. Active contraction produces a further increase by prestressing the surrounding network and recruiting tensile load-bearing paths. The corresponding displacement maps in Fig.~\ref{fig:supp_translation_maps} confirm that the voided network undergoes the largest local translation, whereas spheroid-containing systems constrain motion around the cavity.

For solid-like spheroids, the modulus at $\gamma_m=0.01$ increases systematically with the pre-shear prestress magnitude $|\sigma|$ [Fig.~\ref{fig:single_spheroid_response}(b)]. Consistent with the persistent-interface limit, the solid-like data are well described by Eq.~\eqref{eq:permanent_linker_stiffening}, with $\alpha\simeq1.4$. Here, $G_{\mathrm{pass}}$ is a fitted offset close to the modulus of the matched passive solid-like system, so the fitted excess $G-G_{\mathrm{pass}}$ approximates the reinforcement associated with active loading within this protocol. The exponent is interpreted as an effective small-strain stress-stiffening exponent rather than a universal value, because it can depend on network architecture, boundary conditions, geometry, and the definition of the apparent modulus.

The spatial measurements support the nonlinear-load-recruitment picture underlying this scaling. The prestress snapshots in Fig.~\ref{fig:supp_prestress_snapshots_4.77} show that increasing active contraction extends tensile force chains away from the spheroid. Correspondingly, Fig.~\ref{fig:single_spheroid_response}(c) shows enhanced fiber tension near the spheroid surface that decays into the matrix. The buckling-rich region grows at the same time: the radius $R_{\mathrm b}$ containing $75\%$ of buckled bonds increases from $0$ in the passive system to $10.5$, $12.0$, and $12.5$ across the three active solid-like systems. This coexistence of local buckling and extended tensile transmission is consistent with nonlinear active-inclusion theories in which compression is relaxed by buckling while tension propagates through rope-like force chains~\cite{Ronceray2016}. It also provides a microscopic interpretation of the effective prestress contribution in the theoretical framework: active contraction changes not only the tension of existing fibers but also the extent of the network recruited into load-bearing pathways~\cite{Vahabi2016AxialPrestress,Heidemann2015Elasticity3DNetworks,Arzash2019,Kory2024DiscreteContinuum}.

To separate the effects of inclusion rigidity and interfacial remodeling, we replaced the vertex-model spheroid with a fixed-topology spherical spring network [Fig.~\ref{fig:supp_toy_model_simulations}]. The inclusion contains only stretching interactions and is either floppy, with $z\simeq4$, or rigid, with $z\simeq8$. The external network, inclusion-node positions, and permanent linker connections are matched between the two cases, and no linker turnover or internal rearrangement is allowed. The two inclusion types have similar noncontractile moduli and, after subtraction of their respective baselines, similar active excess moduli at closely matched external-fiber prestress. Because $\chi_L=0$, Eq.~\eqref{eq:path_transmission_factor} gives $P_L=1$, and Eq.~\eqref{eq:turnover_modulus} reduces to the permanent-interface relation in Eq.~\eqref{eq:permanent_linker_stiffening}. The toy-model comparison therefore indicates that inclusion rigidity alone produces only a weak difference under the conditions examined and cannot account for the pronounced solid--fluid contrast in the full model.

This contrast directs attention to the dynamics of the spheroid--matrix interface. Fluid-like spheroids do not follow the same monotonic modulus--prestress relation [Fig.~\ref{fig:single_spheroid_response}(b)], showing that scalar pre-shear prestress is not sufficient to determine the low-strain stiffness. Linkers coupled to solid-like spheroids generally carry larger forces during shear than those coupled to fluid-like spheroids [Fig.~\ref{fig:supp_linker_lability_sigma}(a)], whereas fluid-like spheroids sustain more newly assigned linkers after the initial shear increment [Fig.~\ref{fig:supp_linker_lability_sigma}(b)]. These updates are geometry- or topology-mediated rather than force-threshold rupture events. Each replacement linker is initially force free and can subsequently contract and reload, so renewal need not erase the measured prestress. It does, however, relocate the site and orientation of force application and can reduce the persistence of the interfacial paths that transmit that prestress during shear.

The different interfacial responses are accompanied by distinct cellular organization before shear. Solid-like spheroids concentrate larger maximum shear stresses in the outermost cell layer, whereas fluid-like spheroids exhibit greater cell-shape anisotropy and redistribute shear stress toward the inner layers (Fig.~\ref{fig:SI_stress_anisotropy_layers}). These cellular-scale differences are consistent with a stable stress-bearing boundary in solid-like spheroids and a more deformable, rearranging interface in fluid-like spheroids.

Figure~\ref{fig:single_spheroid_response}(d) directly tests the turnover-modified modulus predicted by Eq.~\eqref{eq:turnover_modulus}. At $\gamma_m=0.01$, we define $\Delta G=G-G_{\mathrm{pass}}$ using a separate matched passive baseline for each spheroid type and fix $\alpha=1.4$ from the solid-like prestress scaling. Taking the logarithm of Eq.~\eqref{eq:turnover_modulus} gives
\begin{equation}
\ln\left(\frac{\Delta G}{|\sigma|^{1.4}}\right)
=a_{\mathrm{type}}-\lambda_{\mathrm{type}}\chi_L.
\label{eq:turnover_linearized_fit}
\end{equation}
Here, the type-dependent intercept represents the prestress-to-stiffness conversion coefficient, and the negative slope gives $\lambda_{\mathrm{type}}$. With the constraint $\lambda_{\mathrm{type}}\geq0$, the solid-like fit reaches $\lambda_{\mathrm{s}}=0$, indicating that the power-law prestress contribution alone describes the solid-like data over the measured range. The fluid-like fit gives $\lambda_{\mathrm{f}}=32.7$, demonstrating a stronger reduction of persistent load transmission with linker renewal.

Substitution of the fitted $\lambda_{\mathrm{f}}$ into Eqs.~\eqref{eq:path_transmission_factor} and \eqref{eq:turnover_modulus} captures the nonmonotonic fluid-like response in Fig.~\ref{fig:single_spheroid_response}(b). Increasing activity raises the stiffening contribution $|\sigma|^{1.4}$, but it also increases linker renewal and reduces the transmission factor $\exp(-\lambda_{\mathrm{f}}\chi_L)$. When the latter reduction dominates, the modulus decreases despite the larger pre-shear prestress. Thus, the combined results support the two regimes anticipated by the theoretical framework: solid-like spheroids follow the persistent-interface relation in Eq.~\eqref{eq:permanent_linker_stiffening}, whereas fluid-like spheroids require the additional state-dependent turnover correction in Eq.~\eqref{eq:turnover_modulus}. The fitted values of $\lambda$ are effective parameters for the present spheroid geometry and shear protocol.

\subsection{Geometry-dependent prestress and stiffening in single-spheroid networks}
\label{subsec:single_spheroid_radius}

Having separated prestress generation from interfacial transmission at fixed spheroid size, we next examine how changing the active-boundary geometry modifies these contributions. The active-linker parameters are fixed while the radius $R$ of a single spheroid is varied. Within the theoretical framework of Sec.~\ref{subsec:toy_model}, increasing $R$ changes the passive baseline through the inclusion volume fraction $\phi$, the generated prestress through the active interfacial area and linker number, and the efficiency of force transmission through the spacing and overlap of mechanically perturbed regions. The radius dependence should therefore not be interpreted as a purely passive volume-fraction effect.

\begin{figure*}
\centering
\includegraphics[width=\linewidth]{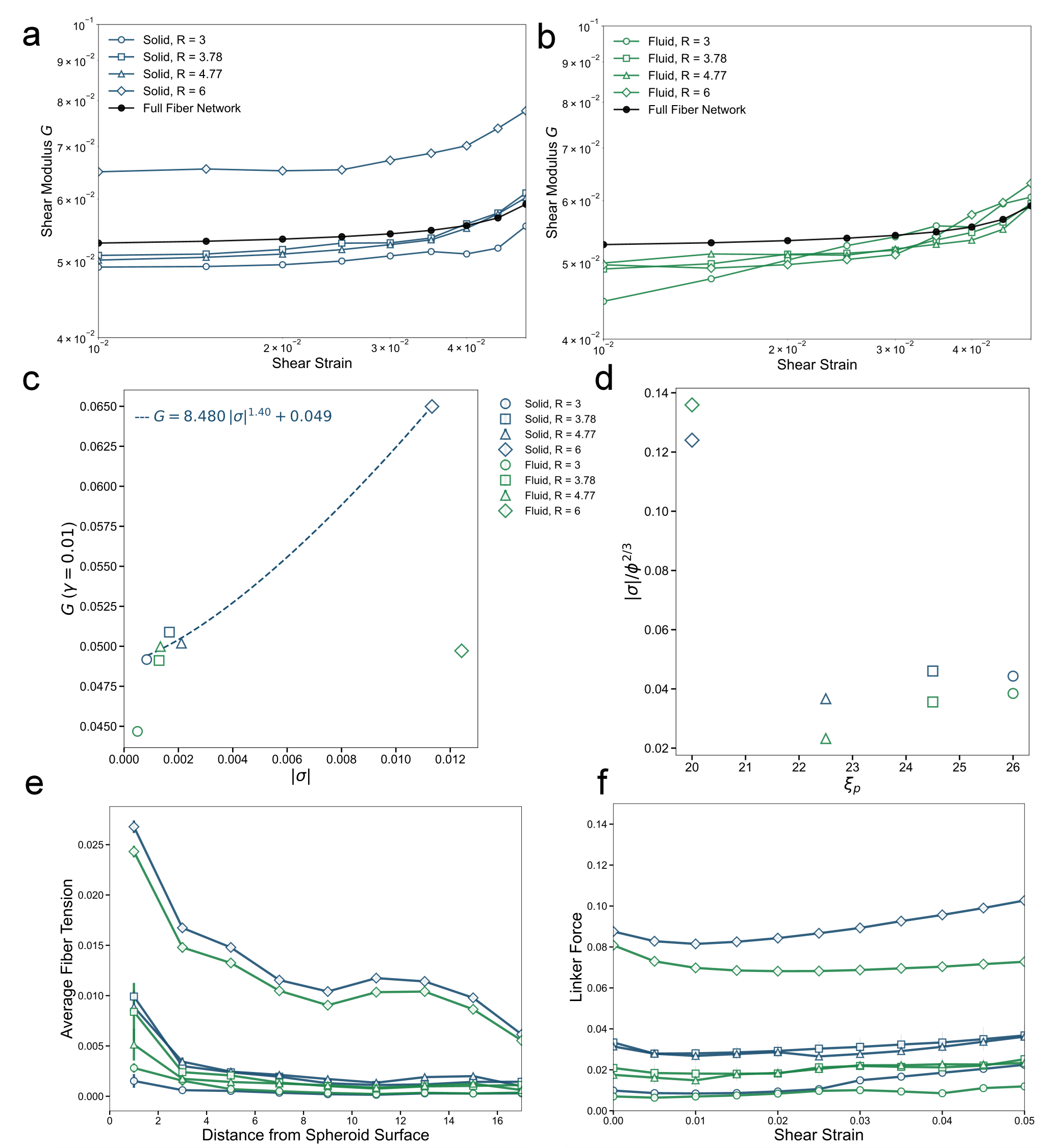}
\caption{\textbf{Geometry-dependent prestress and stiffening in single-spheroid networks.}
(a,b) Apparent shear modulus $G$ as a function of shear strain $\gamma$ for networks containing a single active-linker-coupled (a) solid-like or (b) fluid-like spheroid with radius $R=3$, $3.78$, $4.77$, or $6$. The corresponding pre-shear prestress values are listed in Table~\ref{tab:single_spheroid_prestress}.
(c) Apparent shear modulus at $\gamma_m=0.01$ plotted against the prestress magnitude $|\sigma|$. The dashed curve is a fit to the solid-like data using Eq.~\eqref{eq:permanent_linker_stiffening}, with an effective common offset $G_{\mathrm{pass}}$, giving $\alpha\simeq1.4$ and MSE $1.9\times10^{-7}$.
(d) Surface-area-normalized prestress $|\sigma|/\phi^{2/3}$ plotted against the periodic surface-to-surface spacing $\xi_p=L-2R$.
(e) Average fiber tension as a function of vertical distance from the spheroid surface before shear.
(f) Linker force as a function of shear strain.}
\label{fig:single_radius_increase}
\end{figure*}

The strain-dependent moduli are shown in Figs.~\ref{fig:single_radius_increase}(a) and \ref{fig:single_radius_increase}(b). For solid-like spheroids, the low-strain modulus depends strongly on radius. The systems with $R=3$, $3.78$, and $4.77$ remain close to the intact-network response, whereas the $R=6$ system exhibits pronounced stiffening. This enhancement coincides with a sharp increase in the pre-shear prestress, from $\sigma=-2.11\times10^{-3}$ at $R=4.77$ to $\sigma=-1.13\times10^{-2}$ at $R=6$ (Table~\ref{tab:single_spheroid_prestress}).

Fluid-like spheroids show a much weaker radius dependence. Their moduli remain close to the intact-network response even at $R=6$, despite a prestress of $\sigma=-1.24\times10^{-2}$ that is comparable to that of the corresponding solid-like system. This comparison extends the fixed-radius result of Sec.~\ref{subsec:single_spheroid_prestress}: the magnitude of the pre-shear prestress alone does not determine the modulus when the interface does not preserve coherent load-bearing pathways.

Figure~\ref{fig:single_radius_increase}(c) compares the low-strain modulus with $|\sigma|$. For solid-like spheroids, the radius-scan data are described by the persistent-interface relation in Eq.~\eqref{eq:permanent_linker_stiffening}, with the same effective exponent $\alpha\simeq1.4$ obtained from the fixed-radius prestress scan. Although $C_{\mathrm{pre}}$ can in principle depend on geometry, a single effective coefficient describes the solid-like data over the sampled radii. The passive contribution predicted by Eq.~\eqref{eq:passive_composite_modulus} changes only modestly over the small volume fractions considered here and is represented by an effective fitted offset $G_{\mathrm{pass}}$. By contrast, the fluid-like data require the state-dependent transmission factor in Eq.~\eqref{eq:turnover_modulus}, explaining why comparable values of $|\sigma|$ need not produce comparable moduli.

The growth of prestress with radius can first be understood from the active interfacial area. At fixed linker surface density, the linker number scales approximately as $N_{\mathrm{LS}}\sim R^2$. For a single spheroid in a fixed volume, $\phi\sim R^3$, giving $N_{\mathrm{LS}}\sim\phi^{2/3}$. If the force generated per linker and the efficiency of stress transmission were independent of radius, the resulting prestress would scale approximately as $|\sigma|\sim\phi^{2/3}$. Figure~\ref{fig:single_radius_increase}(d) therefore plots $|\sigma|/\phi^{2/3}$ against the periodic surface-to-surface spacing $\xi_p=L-2R$. For the three smaller radii, the normalized prestress remains of similar magnitude, consistent with an approximately surface-area-controlled regime. For $R=6$, however, $|\sigma|/\phi^{2/3}$ increases sharply for both spheroid types, indicating an additional finite-size amplification mechanism. 

The spatial origin of this amplification is illustrated by the buckling patterns in Fig.~\ref{fig:supp_pretension_snapshots_single}. Active contraction produces a nonlinear region in which compressive stresses are relaxed by fiber buckling while tensile stresses propagate through extended force chains~\cite{Ronceray2016}. We characterize this region using $R_{\mathrm b}$, defined as the radius containing $75\%$ of all buckled bonds. This simulation-defined quantity is not identical to the theoretical crossover radius $R_*$ in Eq.~\eqref{eq:nonlinear_region_radius}, but it provides a measure of the spatial extent of the nonlinear response.

For small and intermediate spheroids, the buckling-rich region remains localized around the inclusion. For the $R=6$ solid-like spheroid, however, $R_{\mathrm b}=15$, approaching the half-box size $L/2=16$. The corresponding buckling-shell thickness, $R_{\mathrm b}-R=9$, is comparable to half of the periodic surface spacing, $\xi_p/2=10$. The fluid-like $R=6$ system exhibits a similarly extended region, with $R_{\mathrm b}=14.2$. The nonlinear regions can therefore interact with their periodic images through extended tensile pathways. This overlap violates the dilute, noninteracting-region assumption underlying Eq.~\eqref{eq:nonlinear_region_modulus} and provides a natural explanation for the enhanced prestress at $R=6$.

The pre-shear fiber-tension profiles confirm that increasing $R$ enhances both the magnitude and spatial extent of tensile loading [Fig.~\ref{fig:single_radius_increase}(e)]. The smaller spheroids produce tension concentrated near the interface, whereas the $R=6$ systems maintain elevated fiber tension over the full distance range examined. Nevertheless, the solid-like and fluid-like $R=6$ systems have broadly comparable pre-shear tension profiles but markedly different moduli. The difference therefore arises primarily during the transmission of this prestressed state under shear.

Solid-like spheroids maintain comparatively stable linker attachment geometries, allowing the linker force to increase after a small initial relaxation. Fluid-like spheroids instead sustain lower and more weakly strain-dependent linker forces [Fig.~\ref{fig:single_radius_increase}(f)] together with more persistent geometry-driven linker renewal [Fig.~\ref{fig:supp_linker_lability_radius}]. Newly assigned linkers can contract and reload, but their changing positions and orientations redistribute traction and interrupt existing force chains. In the language of Eq.~\eqref{eq:path_transmission_factor}, this behavior reduces the effective transmission factor $P_L$. 

The radius-dependent response thus consists of two coupled stages. Increasing $R$ first increases the active interfacial area and extends the prestressed region, with an additional finite-size amplification when the nonlinear region approaches its periodic images. Spheroid fluidity then determines how efficiently this prestress is converted into shear stiffness. Solid-like spheroids preserve coherent interfacial pathways and approximately follow Eq.~\eqref{eq:permanent_linker_stiffening}, whereas fluid-like spheroids require the turnover-modified relation in Eq.~\eqref{eq:turnover_modulus}. This distinction explains why the largest solid-like spheroid exhibits pronounced stiffening while the fluid-like spheroid remains close to the intact-network response despite generating a comparable pre-shear prestress.

\subsection{Multiple-spheroid boundary-density effects on network stiffening}
\label{subsec:multiple_spheroid_boundary_density}

The preceding sections established how active prestress and interfacial persistence control stiffening at fixed spheroid geometry, and how changing the radius modifies prestress generation around a single spheroid. We now fix the spheroid radius at $R=3$ and increase the number of spheroids $N_s$ from 1 to 8. At fixed $R$, the inclusion volume fraction $\phi$, total spheroid--matrix interfacial area, and total linker number all increase in proportion to $N_s$, while the average surface-to-surface spacing $\xi_p$ decreases. Within the framework of Sec.~\ref{subsec:toy_model}, changing $N_s$ therefore modifies the passive baseline through $\phi$, the magnitude of active loading through the measured prestress $|\sigma|$, and the prestress-to-stiffness conversion through the spatial organization and persistence of interfacial load paths.

\begin{figure*}
\centering
\includegraphics[width=\linewidth]{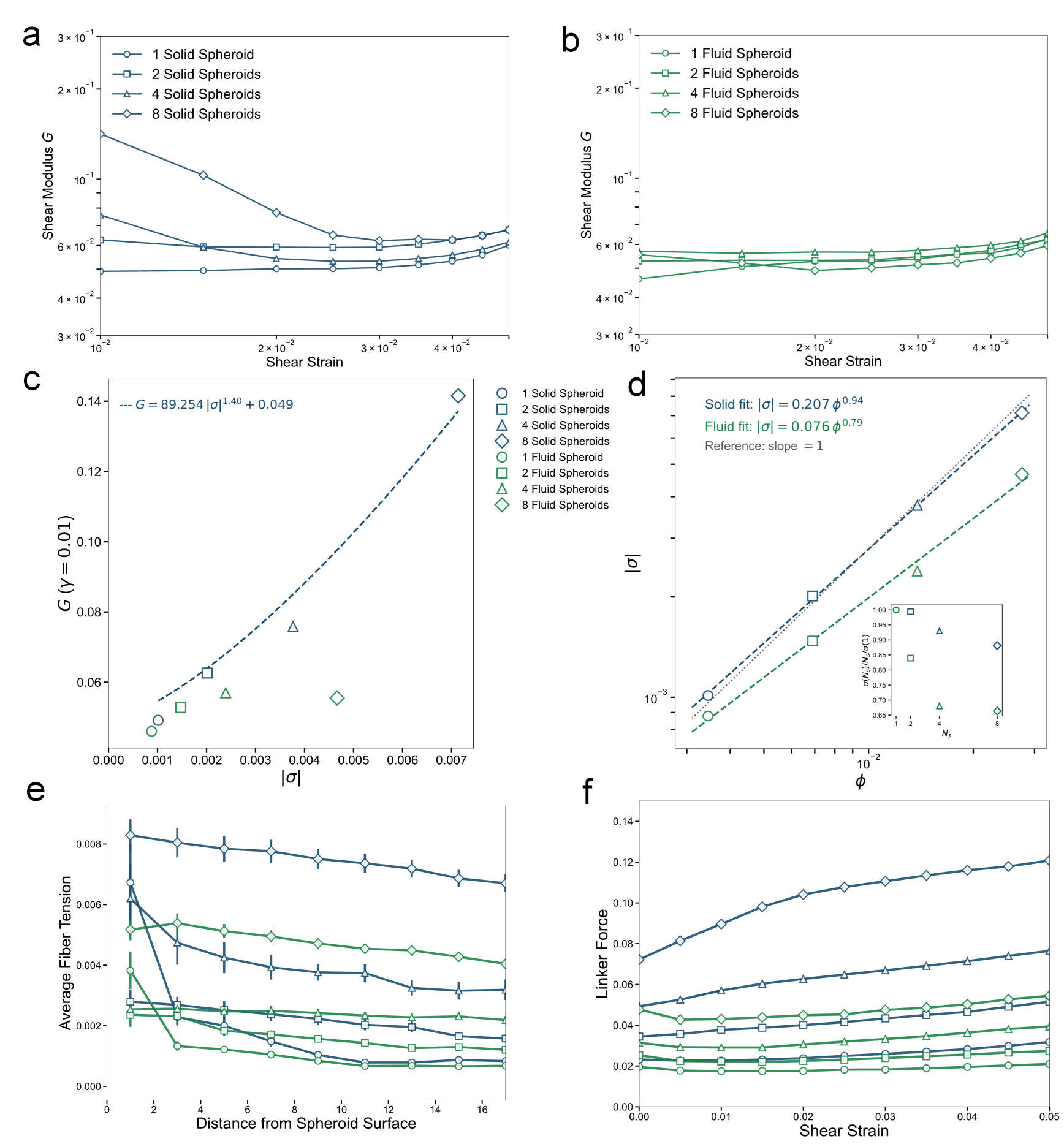}
\caption{\textbf{Collective prestress generation and boundary-density-dependent stiffening in multiple-spheroid networks.}
(a,b) Apparent shear modulus $G$ as a function of shear strain $\gamma$ for networks containing (a) solid-like or (b) fluid-like spheroids of radius $R=3$, with the spheroid number $N_s$ varied from 1 to 8.
(c) Apparent shear modulus at $\gamma_m=0.01$ plotted against the prestress magnitude $|\sigma|$. The dashed curve is a fit to the solid-like data using Eq.~\eqref{eq:permanent_linker_stiffening}, with an effective common offset $G_{\mathrm{pass}}$, giving $\alpha\simeq1.4$ and MSE $3.4\times10^{-5}$.
(d) Prestress magnitude $|\sigma|$ plotted against the spheroid volume fraction $\phi$ on logarithmic axes. The dotted line with slope $\beta=1$ denotes additive prestress generation. Effective fits give $|\sigma|\sim\phi^{0.94}$ for solid-like spheroids and $|\sigma|\sim\phi^{0.79}$ for fluid-like spheroids. The inset shows the prestress additivity ratio $\mathcal{E}_{\sigma}$ defined in Eq.~\eqref{eq:prestress_additivity_ratio}, normalized separately using the corresponding single-spheroid prestress for each spheroid type.
(e) Average fiber tension as a function of distance from the spheroid surface before shear.
(f) Linker force as a function of shear strain.}
\label{fig:multiple_spheroid_mechanics}
\end{figure*}

The solid-like systems exhibit a strong number dependence at small strain. At $\gamma_m=0.01$, the apparent shear modulus increases from approximately $0.049$ for one spheroid to $0.142$ for eight spheroids, corresponding to an almost threefold enhancement [Fig.~\ref{fig:multiple_spheroid_mechanics}(a)]. This enhancement is strongest at low strain and decreases as the applied strain increases. By contrast, the fluid-like systems remain within a narrower modulus range and show only a weak, nonmonotonic dependence on spheroid number [Fig.~\ref{fig:multiple_spheroid_mechanics}(b)]. The different responses indicate that reinforcement depends not only on the amount of embedded material but also on how the active interfaces generate and transmit stress.

The leading passive-inclusion contribution in Eq.~\eqref{eq:passive_composite_modulus} depends on the total volume fraction. A fixed-volume comparison therefore provides a useful test of whether $\phi$ alone describes the response. The pairs $2R3$ and $1R3.78$, $4R3$ and $1R4.77$, and $8R3$ and $1R6$ have approximately equal total spheroid volumes, yet distributing this volume among several smaller solid-like spheroids produces a higher modulus [Fig.~\ref{fig:supp_aggregated_vs_dispersed}]. At fixed total volume, dispersion increases the total active interfacial area and changes the spacing between active boundaries. The observed difference therefore lies beyond the leading passive volume-fraction contribution and demonstrates that active-boundary area and spatial organization also affect the mechanical response. 

We next quantify how the prestress accumulates as additional fixed-radius spheroids are introduced. Because both the active interfacial area and $\phi$ are proportional to $N_s$, independent and additive stress generation would give $|\sigma|\propto N_s\propto\phi$. Figure~\ref{fig:multiple_spheroid_mechanics}(d) shows that the solid-like systems remain close to this additive limit, with an effective scaling $|\sigma|\sim\phi^{0.94}$. The fluid-like systems exhibit a more sublinear dependence, $|\sigma|\sim\phi^{0.79}$. 

To remove the trivial increase in the number of active boundaries, we define the prestress additivity ratio
\begin{equation}
\mathcal{E}_{\sigma}(N_s)
=
\frac{|\sigma(N_s)|}
{N_s|\sigma(1)|}.
\label{eq:prestress_additivity_ratio}
\end{equation}
The additive limit corresponds to $\mathcal{E}_{\sigma}=1$. As shown in the inset of Fig.~\ref{fig:multiple_spheroid_mechanics}(d), $\mathcal{E}_{\sigma}$ decreases only modestly to approximately $0.88$ for eight solid-like spheroids, whereas it falls to approximately $0.66$ for eight fluid-like spheroids. Thus, solid-like boundaries retain most of their single-spheroid prestress contribution as $N_s$ increases, whereas the average prestress contribution per fluid-like boundary progressively decreases.

The spatial organization accompanying this prestress accumulation is shown in Fig.~\ref{fig:supp_pretension_snapshots}. With one or two spheroids, the tensile and buckling-rich regions remain comparatively localized. As $N_s$ increases, the nonlinear regions surrounding neighboring spheroids approach and overlap, while tensile pathways extend through a larger fraction of the network. Consistently, the eight-spheroid solid-like system maintains elevated average fiber tension over the full distance range examined, whereas the tension produced by fewer spheroids decays more strongly away from their surfaces [Fig.~\ref{fig:multiple_spheroid_mechanics}(e)].

This overlap marks a departure from the dilute, noninteracting-region assumption underlying Eq.~\eqref{eq:nonlinear_region_modulus}. It does not produce a super-additive scalar prestress, because $\mathcal{E}_{\sigma}$ remains below unity. Instead, the additional solid-like boundaries broaden and connect the prestressed regions while retaining most of their additive contribution to the global stress. The resulting spatial connectivity can also modify the geometry-dependent conversion coefficient $C_{\mathrm{pre}}$ by allowing prestress to be transmitted through a more extended set of load-bearing pathways.

The fluid-like systems also develop broader tension and buckling distributions as $N_s$ increases, but their prestress accumulation is more subadditive and their modulus remains weakly number dependent. Spatial overlap of the mechanically perturbed regions is therefore not sufficient for collective stiffening. At the measurement strain $\gamma_m=0.01$, Eq.~\eqref{eq:turnover_modulus} combines the initial prestress magnitude $|\sigma|$ with the persistence of the interfacial load paths over the same low-strain interval.

Linker forces increase with spheroid number for both spheroid types, but solid-like spheroids sustain larger forces and a stronger strain-dependent increase [Fig.~\ref{fig:multiple_spheroid_mechanics}(f)]. Fluid-like spheroids also exhibit larger linker-creation counts during the initial strain increments [Fig.~\ref{fig:supp_linker_lability_number}]. These early renewal events contribute directly to the cumulative count $N_{\mathrm{new}}(0,\gamma_m)$ and therefore provide a measure of the interfacial lability entering $\chi_L(\gamma_m)$ in Eq.~\eqref{eq:normalized_linker_turnover}. Their larger values for fluid-like spheroids indicate less persistent load transmission over the strain interval used to evaluate the modulus and hence a smaller effective $P_L$ in Eq.~\eqref{eq:path_transmission_factor}.

For solid-like spheroids, the nearly additive growth of prestress is efficiently converted into low-strain stiffness. The data in Fig.~\ref{fig:multiple_spheroid_mechanics}(c) are described by the persistent-interface relation in Eq.~\eqref{eq:permanent_linker_stiffening}, with the same effective exponent $\alpha\simeq1.4$ found in the preceding sections. Although $G_{\mathrm{pass}}$ and $C_{\mathrm{pre}}$ can in principle depend on $N_s$ and spatial organization, a single effective coefficient and offset describe the solid-like data over the sampled range. Combining $|\sigma|\sim\phi^{0.94}$ with $\alpha\simeq1.4$ gives the approximate effective dependence $G-G_{\mathrm{pass}}\sim\phi^{1.32}$. 

Fluid-like spheroids exhibit both subadditive prestress accumulation and greater interfacial lability within the initial strain window. Their response therefore requires the full relation in Eq.~\eqref{eq:turnover_modulus}: increasing $N_s$ raises the total prestress, while the larger $\chi_L$ reduces the transmission factor $P_L$. Additional spheroids thus reinforce the network by increasing the active-boundary density and connecting mechanically perturbed regions, whereas spheroid fluidity controls how persistently these prestressed pathways transmit load at $\gamma_m=0.01$. This distinction explains the strong collective stiffening of solid-like systems and the much weaker number dependence of fluid-like systems. The progressive reduction of this boundary-density-dependent reinforcement with strain motivates the large-strain analysis below.

\subsection{Large-strain mechanics of spheroid-embedded fiber networks}
\label{subsec:large_strain_mechanics}

The preceding sections showed that spheroid rigidity, boundary density, active prestress, and interfacial persistence strongly affect the modulus at the measurement strain $\gamma_m=0.01$. We now examine how these contributions evolve after the surrounding fiber network enters its nonlinear strain-stiffening regime. Figure~\ref{fig:large_strain_converge} compares networks containing eight solid-like or fluid-like spheroids of radius $R=3$ with an intact fiber network and a matched eight-void reference. At small strain, the solid-like spheroids produce substantial reinforcement, whereas the fluid-like spheroids produce a weaker response. As strain increases, however, both spheroid-containing systems approach the eight-void response while remaining softer than the intact network. The same convergence occurs for $N_s=1$, $2$, $4$, and $8$, for which $G_{\mathrm{spheroid}}/G_{\mathrm{void}}$ approaches unity at large strain [Fig.~\ref{fig:supp_voided_fiber_network}]. Because the reported $G$ is an apparent modulus, the initial decrease of the solid-like curve for $\gamma\lesssim0.03$ may partly reflect the diminishing contribution of a pre-existing shear-stress offset and should not by itself be interpreted as linker unloading or interfacial failure.

In the low-strain framework, the excess modulus relative to the matched voided network contains a passive contribution, $G_{\mathrm{pass}}-G_{\mathrm{void}}$, described by Eq.~\eqref{eq:passive_composite_modulus}, and a prestress-dependent interfacial contribution, $G-G_{\mathrm{pass}}$, described by Eq.~\eqref{eq:turnover_modulus}. These relations were formulated for the low-strain regime using the pre-shear prestress and the interfacial persistence over the corresponding measurement interval. They are therefore not expected to remain quantitatively valid at large strain. As shear-induced fiber tension and stretching stiffness increase, the combined inclusion and interfacial correction becomes small relative to the rapidly increasing modulus of the residual voided backbone. The convergence toward $G_{\mathrm{void}}$ does not require the passive and active contributions to vanish separately; it shows that their net contribution becomes mechanically subdominant.

The force and energy measurements support this crossover. Fiber tension increases rapidly with strain, and the spheroid-containing systems approach the matched voided-network response [Fig.~\ref{fig:large_strain_forces}(a)]. Meanwhile, the linker force remains finite and continues to increase, with solid-like spheroids sustaining larger forces than fluid-like spheroids [Fig.~\ref{fig:large_strain_forces}(b)]. Thus, a linker can remain loaded while contributing relatively little to the incremental stiffness. Because the model contains no force-dependent linker-rupture rule, the convergence cannot be attributed to linker failure. At the same time, differences in nonaffinity and elastic energy among the spheroid-containing systems progressively diminish, while the stretching energy grows much more strongly than the bending energy [Fig.~\ref{fig:supp_nonaffinity_energy_partition}]. These trends are consistent with a response increasingly controlled by stretching of the residual fiber backbone.

The system therefore crosses over from a low-strain regime governed by active prestress and persistent interfacial transmission to a large-strain regime governed primarily by the remaining fiber network. Under the present linker density and geometry, the loaded interface establishes the prestressed initial state but does not restore the continuous fiber pathways removed when the cavities are created. Consequently, the large-strain response is controlled by the matched voided backbone rather than by the intact fiber network.

\section{Discussion}

This study establishes a hierarchical framework for the mechanics of active spheroid--fiber composites. Relative to a matched voided network, passive spheroid--matrix coupling provides modest reinforcement by suppressing cavity-like nonaffine relaxation. Active linker contraction produces a larger contribution by prestressing the nonlinear fiber network and recruiting tensile load-bearing pathways. The low-strain response is summarized by Eq.~\eqref{eq:turnover_modulus}, in which the active excess modulus is controlled jointly by the prestress magnitude $|\sigma|$, the conversion coefficient $C_{\mathrm{pre}}$, and the path-transmission factor $P_L=\exp(-\lambda\chi_L)$. These quantities are not independent: the measured prestress already contains the effects of linker activity, spheroid number, and active interfacial area, whereas $C_{\mathrm{pre}}$ and $\lambda$ can depend on spheroid state, geometry, and the spatial organization of the mechanically perturbed regions. Over the sampled ranges, the solid-like data in the prestress, radius, and spheroid-number scans are approximately described by an effective exponent $\alpha\simeq1.4$, although the fitted baseline and conversion coefficient can vary with geometry. This exponent is close to the value $3/2$ suggested by the simple nonlinear-region volume estimate based on active-inclusion theory, but this agreement should not be interpreted as a universal prediction~\cite{Ronceray2019}. The nonmonotonic fluid-like response demonstrates that scalar prestress alone is insufficient when interfacial transmission is not persistent.

The fixed-topology spring-network toy model separates the direct effect of inclusion rigidity from interfacial rearrangement. With identical external networks, inclusion-node positions, linker endpoints, and permanent linkers, spherical inclusions with $z\simeq4$ and $z\simeq8$ produce similar passive moduli and similar active excess moduli at approximately matched external-fiber prestress [Fig.~\ref{fig:supp_toy_model_simulations}]. Thus, although inclusion rigidity can enter both the passive response and $C_{\mathrm{pre}}$, its direct effect is comparatively weak under the present fixed-interface conditions. The larger solid--fluid difference in the vertex-model system therefore points primarily to differences in the organization and persistence of interfacial force transmission. Consistently, the constrained turnover fits give $\lambda_{\mathrm{s}}=0$ for solid-like spheroids and $\lambda_{\mathrm{f}}=32.7$ for fluid-like spheroids [Fig.~\ref{fig:single_spheroid_response}(d)]. The result $\lambda_{\mathrm{s}}=0$ does not imply an absence of linker updates; it indicates that, after accounting for the $|\sigma|^{1.4}$ contribution, no additional turnover-induced reduction is resolved for the solid-like data. For fluid-like spheroids, the decrease of $\exp(-\lambda_{\mathrm{f}}\chi_L)$ can outweigh the increase of $|\sigma|^{1.4}$ and thereby account for the nonmonotonic modulus--prestress relation. Because renewed linkers can contract and reload, $\chi_L$ measures pathway relocation rather than the fraction of prestress permanently lost. Moreover, it does not record linker force, orientation, or subsequent loading history. The fitted values of $\lambda$ are therefore effective parameters for the present geometry and loading protocol rather than universal material constants. The larger outer-layer cellular shear stresses in solid-like spheroids and greater cellular anisotropy in fluid-like spheroids provide an additional cellular-scale signature of these distinct interfacial states [Fig.~\ref{fig:SI_stress_anisotropy_layers}].

Spheroid radius and number determine how active stress is generated and organized throughout the matrix. Increasing the radius increases the inclusion volume, active interfacial area, and spatial extent of the nonlinear prestressed region. For the $R=6$ single-spheroid systems, this region approaches the periodic images of the same inclusion, producing a finite-size amplification of prestress. Increasing $N_s$ at fixed $R$, in contrast, causes nonlinear regions generated by distinct spheroids to approach and overlap, producing a collective boundary-density effect. Both mechanisms violate the dilute, noninteracting-region assumption underlying Eq.~\eqref{eq:nonlinear_region_modulus}, but their geometrical origins are different. The additivity ratio in Eq.~\eqref{eq:prestress_additivity_ratio} remains below unity, showing that overlap does not generate super-additive scalar prestress. Instead, multiple solid-like boundaries broaden and connect persistent tensile pathways while retaining most of their additive stress contribution. Over the sampled range, the effective scalings are $|\sigma|\sim\phi^{0.94}$ for solid-like spheroids and $|\sigma|\sim\phi^{0.79}$ for fluid-like spheroids. Combining the solid-like scaling with $\alpha\simeq1.4$ gives the approximate relation $G-G_{\mathrm{pass}}\sim\phi^{1.32}$, provided that the effective baseline and conversion coefficient vary weakly over this range. These exponents describe the present data and should not be interpreted as universal scaling laws. The fixed-volume comparisons further establish that $\phi$ alone is insufficient, although they do not separate the increased active interfacial area from the accompanying change in inter-spheroid spacing and pathway connectivity. Because these geometrical effects already influence the measured $|\sigma|$ and $C_{\mathrm{pre}}$, no additional simple factor of $\phi$ or $N_s$ should be multiplied into the active term.

The influence of the active interface becomes progressively less important as the fiber network strain stiffens. Equations~\eqref{eq:passive_composite_modulus} and \eqref{eq:turnover_modulus} describe the low-strain response and are not intended as quantitative large-strain constitutive relations. At large strain, the moduli of both solid-like and fluid-like systems approach those of the corresponding voided networks for all spheroid numbers examined [Figs.~\ref{fig:large_strain_converge} and \ref{fig:supp_voided_fiber_network}]. The linkers nevertheless remain loaded, and their forces continue to increase [Fig.~\ref{fig:large_strain_forces}(b)]. The convergence therefore does not imply that $P_L$ vanishes, that the linkers unload, or that the passive and active interfacial contributions disappear separately. Rather, their net incremental contribution becomes small relative to the rapidly increasing stretching stiffness of the residual fiber backbone. The simultaneous reduction of differences in nonaffinity and elastic energy among spheroid-containing systems, together with the increasing dominance of stretching over bending energy, supports this interpretation [Fig.~\ref{fig:supp_nonaffinity_energy_partition}]. Because the model contains no force-dependent linker-rupture rule, the crossover reflects a change in the dominant load-bearing mechanism rather than interfacial failure.

The observed reinforcement is substantially larger than expected from passive inclusion volume alone. As a rough benchmark, the conventional rigid-inclusion estimate summarized by Shivers, Feng, and MacKintosh predicts only an approximately $0.7\%$--$5.7\%$ modulus increase over the single-spheroid volume fractions studied here, far below the enhancement of the active solid-like systems~\cite{MacKintosh2025}. Near a rigidity transition, rigid inclusions can produce stronger reinforcement by limiting the network correlation length through their spacing~\cite{MacKintosh2025}; however, our passive controls and fixed-topology toy model indicate that passive inclusion rigidity alone cannot account for the strong state-dependent reinforcement observed here. The magnitude of active reinforcement should also depend on fiber bending rigidity: increasing $K_B$ raises the low-strain elastic baseline and reduces sensitivity to prestress, whereas decreasing $K_B$ moves the network toward the central-force limit, where prestress-induced stiffening becomes increasingly important~\cite{Mao2010,das2012redundancy,Sharma2016-kf,Vahabi2016AxialPrestress,Arzash2019}. Thus, proximity to network rigidity provides an additional control parameter that should tune how efficiently cell-generated activity is converted into tissue-scale stiffness.

The contraction--relaxation test further shows that scalar prestress does not fully characterize the low-strain response. Reducing the linker rest-length decrement per FIRE iteration lowers the modulus and linker force of solid-like systems despite only modest changes in pre-shear prestress [Fig.~\ref{fig:SI_linker_contraction}]. Because contraction and mechanical relaxation occur concurrently, changing the update decrement can alter the balance between tensile-path recruitment and nonaffine stress redistribution, producing different load-bearing structures at similar $|\sigma|$. Fluid-like systems show no systematic protocol dependence, consistent with interfacial rearrangements that continually redistribute traction. At larger strains, these differences diminish as the residual fiber backbone becomes dominant.

The model also suggests a direct experimental test of the role of activity in multi-spheroid mechanics. Reducing actomyosin contractility within the spheroids, for example through treatment with blebbistatin, should decrease the active prestress transmitted to the surrounding ECM and consequently reduce the recruitment of tensile load-bearing pathways. For multiple spheroids at fixed size, number, and spacing, we therefore predict that increasing blebbistatin concentration will progressively suppress the collective stiffening of the shared matrix, with the mechanical response approaching the passive limit as contractility is reduced. This effect should be particularly pronounced for solid-like spheroids, for which persistent cell–ECM coupling allows active prestress to be efficiently converted into matrix stiffening, whereas the response of fluid-like spheroids is already weakened by interfacial rearrangements. More generally, independently varying contractility and collective cell fluidity would provide an experimental means of testing the central prediction of the model: tissue-scale mechanics is controlled not simply by the magnitude of cell-generated forces, but by the interplay between active prestress and the persistence of the pathways that transmit those forces through the ECM.

One important future direction is to extend the present framework by incorporating plastic remodeling of the fiber network. Collagen networks can retain cell-induced densification, alignment, and residual deformation through stress-activated fiber sliding and merging, deformation-induced interfiber bonding, and permanent fiber elongation~\cite{BAN2018450,kim2017stress}. Including these mechanisms would make it possible to determine how the relatively persistent traction patterns of solid-like spheroids and the spatially redistributed traction patterns of fluid-like spheroids are encoded as lasting structural and mechanical memory in the ECM. Traction-release and loading--unloading simulations could further quantify residual strain, fiber organization, and stiffness as functions of spheroid fluidity and loading history.

Such extensions may also help connect the model to recent experiments on multiple cell clusters in fibrous matrices. In one setting, cells interacting with a deformable fiber network spontaneously assemble into multiple spheroids that subsequently interact, exchange cells, and merge, while experiments with pairs of contractile fibroblast spheroids demonstrate long-range mechanical communication through the formation of aligned matrix regions between neighboring spheroids~\cite{Sharma2025Mechanical,Chen2022Glycosaminoglycans}. Although the present simulations exhibit overlapping mechanically perturbed regions and collective stiffening with increasing spheroid number, they do not produce similarly localized inter-spheroid mechanical bridges. This difference suggests that matrix remodeling, fiber alignment, and/or cellular migration may provide additional ingredients that focus distributed force transmission into persistent structures connecting neighboring cell clusters. Incorporating these processes would therefore allow the model to address not only how multiple cell clusters collectively alter tissue-scale mechanics, but also how their mechanical interactions reshape the ECM and ultimately one another.

A complementary direction is to connect the quasi-static mechanical response studied here with the dynamical remodeling regime examined in our previous work~\cite{zhang2025}. This work showed that fluid-like spheroids produce greater cumulative fiber displacement and densification over long remodeling times, whereas the present results reveal stronger immediate reinforcement by solid-like spheroids. To bridge these regimes, active linker rest-length evolution could be incorporated throughout a prolonged overdamped pre-shear stage, allowing cellular rearrangements, linker contraction and renewal, and matrix reorganization to develop concurrently before shear. This extension would reveal how accumulated remodeling modifies prestress, fiber organization, and subsequent stiffness, and whether it changes the relative reinforcement produced by solid-like and fluid-like spheroids. Based on recent work, additional extensions include multi-scale aspects of tissues down to the chromatin scale as well as the learning capability of tissues to ultimately formulate a more comprehensive theoretical framework~\cite{zhang2026human,ameen2026training}. 


\section*{Acknowledgments}
Tao Zhang acknowledges financial support from the NSFC/China via award 22303051. The computations in this paper were run on the \(\pi\) 2.0 and the Siyuan-1 cluster supported by the Center for High Performance Computing at Shanghai Jiao Tong University. JMS acknowledges financial support from the National Science Foundation under Grant PoLS-2412961.

\section*{Data Availability}
The data and code required to reproduce our analyses are available on Zenodo at 10.5281/zenodo.22144019.

\bibliography{manuscript}

\clearpage
\onecolumngrid

\begin{center}
{\large\bfseries Supplemental Material for}\\[0.5em]
{\Large\bfseries \papertitle\par}

\vspace{1.2em}

{\large
Liyang Wang,$^{1}$
J. M. Schwarz,$^{2,3,*}$
and Tao Zhang$^{1,\dagger}$\par}

\vspace{0.7em}

{\small
$^{1}$School of Chemistry and Chemical Engineering,
Shanghai Jiao Tong University, Shanghai 200240, China\\
$^{2}$Physics Department, Syracuse University,
Syracuse, New York 13244, USA\\
$^{3}$Indian Creek Farm, Ithaca, New York 14850, USA\\[0.4em]
$^{*}$jmschw02@syr.edu\\
$^{\dagger}$zhangtao.scholar@sjtu.edu.cn
}
\end{center}

\vspace{1.5em}

\setcounter{page}{1}
\renewcommand{\thepage}{S\arabic{page}}

\setcounter{figure}{0}
\renewcommand{\thefigure}{S\arabic{figure}}

\clearpage
\begin{figure}[H]
\centering
\includegraphics[width=0.95\linewidth]{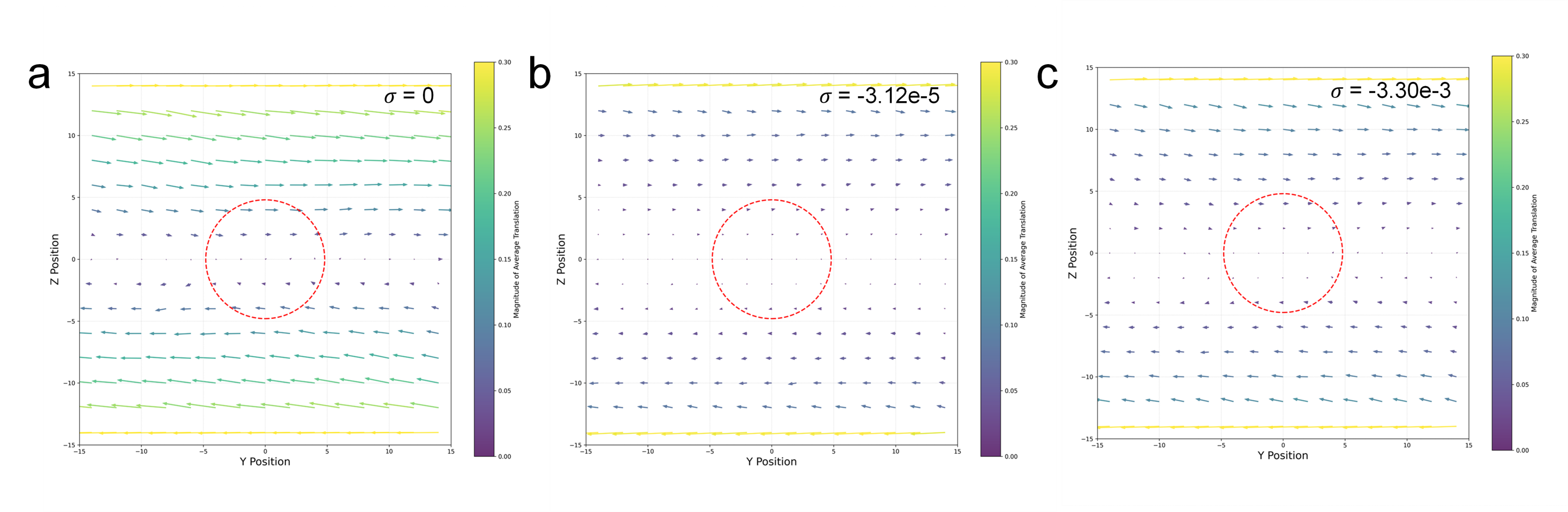}
\caption{Visualized average fiber translation of \textbf{(a)} voided fiber network \textbf{(b)} solid spheroid embedded with passive linker \textbf{(c)} solid spheroid embedded with active linker at 3\% global strain.  The \(\sigma\) notation is prestress exerted by spheroid at 0\% strain.}
\label{fig:supp_translation_maps}
\end{figure}

\begin{figure}[H]
\centering
\includegraphics[width=0.8\linewidth]{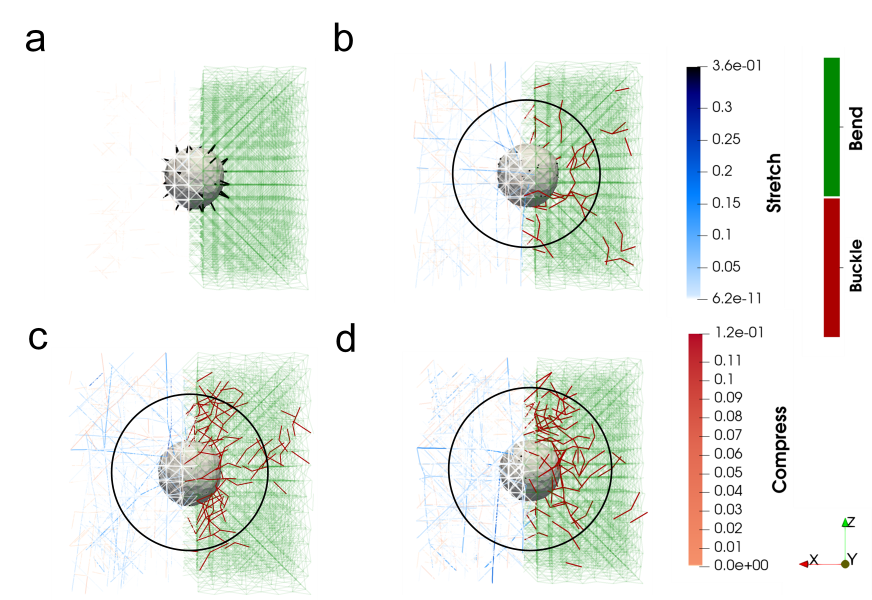}
\caption{Snapshots of fiber stress and buckle distribution in networks containing (a--d) a solid-like spheroid at 0\% strain, with prestress values of \(-3.3\times10^{-5}\), \(-2.11\times10^{-3}\), \(-2.74\times10^{-3}\), and \(-3.12\times10^{-3}\), respectively. The left half of each box shows fiber stretching and compression forces, while the right half shows buckled or bent fibers. A bond is considered buckled if the angle of its constituent bonds is smaller than \(150^\circ\). The black circles contain 75\% buckled bonds, with radii 0.0, 10.5, 12.0, and 12.5, respectively. Color intensity represents the magnitude of fiber tension, ranging from low (light) to high (dark).}
\label{fig:supp_prestress_snapshots_4.77}
\end{figure}

\begin{figure}[H]
\centering
\includegraphics[width=1\linewidth]{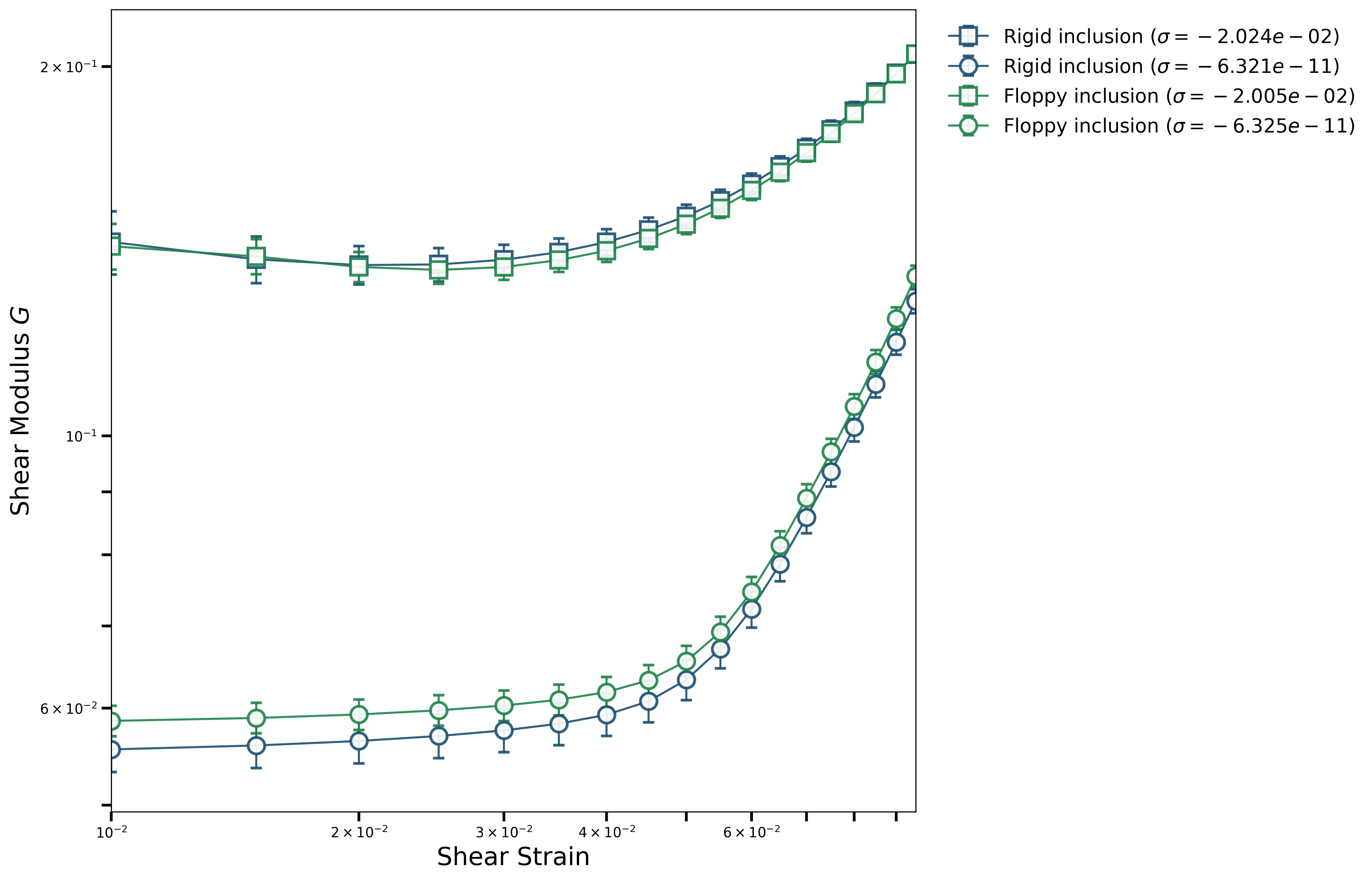}
\caption{\textbf{Prestress-mediated stiffening in the fixed-topology spring-network toy model.}
Apparent shear modulus $G$ as a function of shear strain $\gamma$ for spherical spring-network inclusions with average connectivity $z\simeq4$ (floppy) or $z\simeq8$ (rigid). For each inclusion type, one noncontractile condition and one active condition with final linker rest length $l_{0,\mathrm{L}}^{\mathrm{final}}=0.2$ are shown. The inclusion networks (\(\phi = 2.76\% \)) are non-phantom FCC networks containing stretching interactions only. The rigid and floppy systems use identical node positions, external voided fiber networks, and permanent spheroid--matrix linkers. The values of $\sigma$ in the legend denote the pre-shear prestress calculated from the external fiber bonds only. During shear, the inclusion topology, linker rest lengths, and linker endpoints remain fixed, with no linker rupture, renewal, or reassignment. Curves and error bars represent the mean and standard error over paired network realizations.}
\label{fig:supp_toy_model_simulations}
\end{figure}

\begin{figure}[H]
\centering
\includegraphics[width=0.8\linewidth]{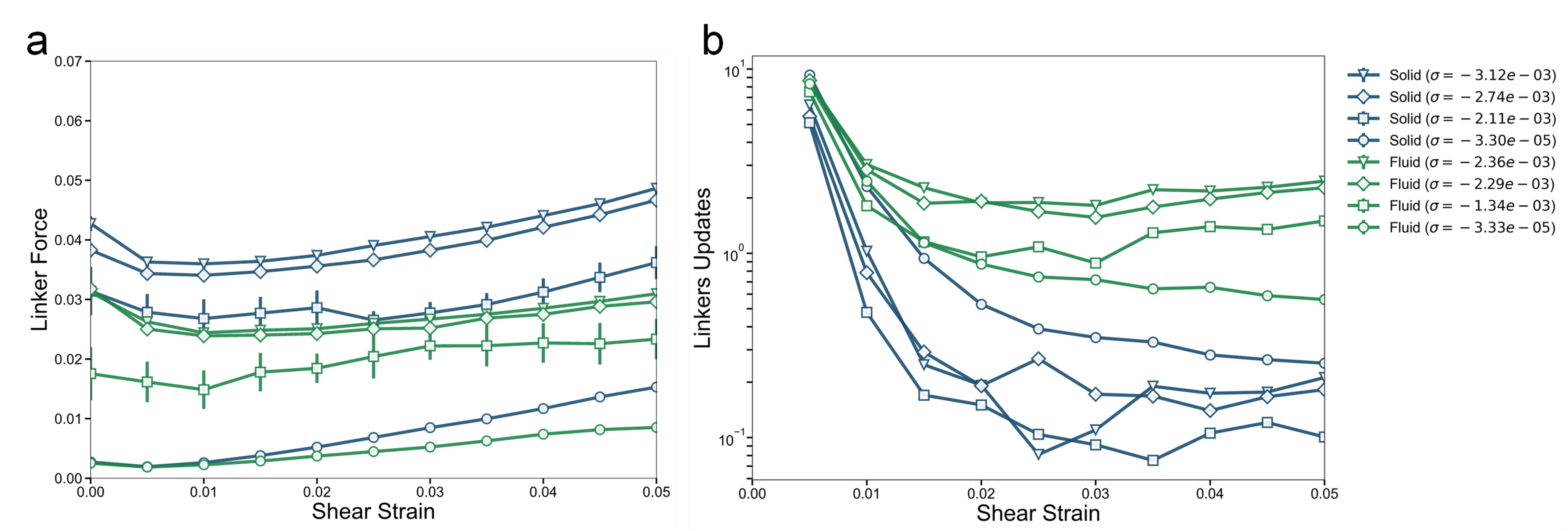}
\caption{Comparison of linker force and linker-update activity between solid-like and fluid-like spheroids.  
(a) Linker force as a function of shear strain. The spheroid radius is $4.77$ and the simulation box size is $32$.
(b)The update count measures newly assigned linkers following geometry- or topology-associated removal events; it is not a count of force-induced rupture. Fluid-like spheroids exhibit higher lability.}
\label{fig:supp_linker_lability_sigma}
\end{figure}

\begin{figure}[H]
\centering
\includegraphics[width=0.95\textwidth]{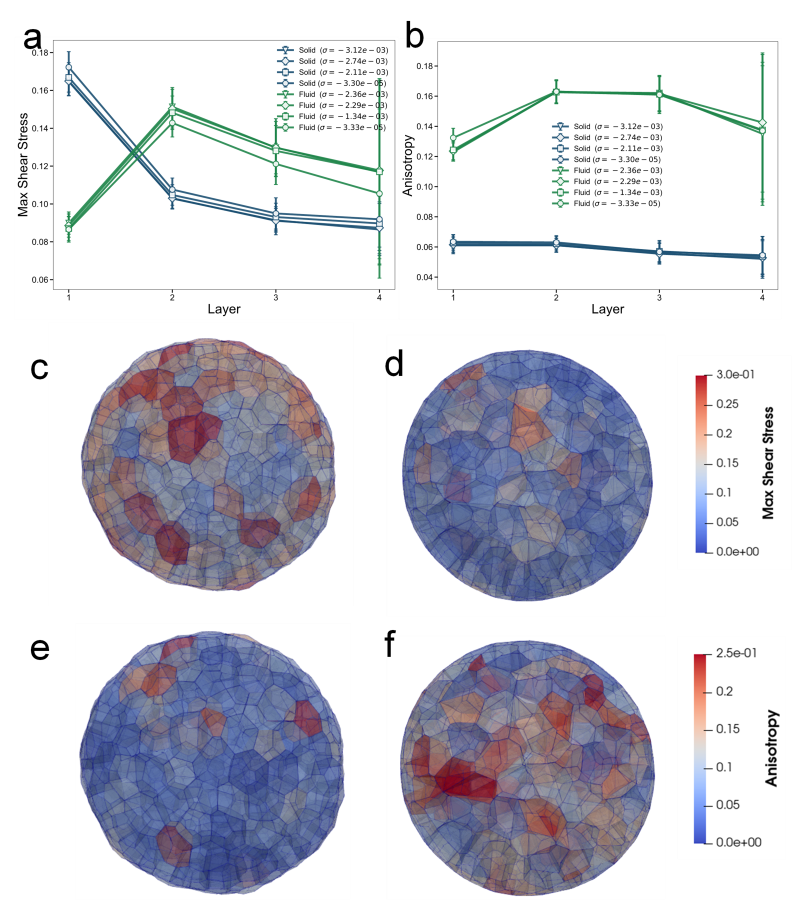}
\caption{
\textbf{Pre-shear layer dependence of cellular stress and shape anisotropy within a single spheroid.}
(a) Maximum shear stress, defined from the largest and smallest principal stresses of the cellular stress tensor as $\tau_{\max}=(\sigma_{\max}-\sigma_{\min})/2$, and (b) cell-shape anisotropy as functions of cell layer for single solid-like and fluid-like spheroids of radius $R=4.77$. Layer 1 denotes the outermost boundary-cell layer, and the layer number increases toward the spheroid center. The anisotropy is calculated from the shape-tensor eigenvalues $\lambda_1\leq\lambda_2\leq\lambda_3$ as $\Delta=\left[\left(\lambda_3-\frac{1}{2}(\lambda_1+\lambda_2)\right)^2+\frac{3}{4}(\lambda_2-\lambda_1)^2\right]/(\lambda_1+\lambda_2+\lambda_3)^2$. The values of $\sigma$ in the legends denote the pre-shear fiber prestress. (c,d) Snapshots of the maximum cellular shear stress for a solid-like spheroid with $\sigma=-2.74\times10^{-3}$ and a fluid-like spheroid with $\sigma=-2.29\times10^{-3}$, respectively. (e,f) Corresponding snapshots of cell-shape anisotropy for the same solid-like and fluid-like systems. All quantities are evaluated at $\gamma=0$ after active contraction and mechanical relaxation but before shear is applied.
}
\label{fig:SI_stress_anisotropy_layers}
\end{figure}

\begin{figure}[H]
\centering
\includegraphics[width=1\linewidth]{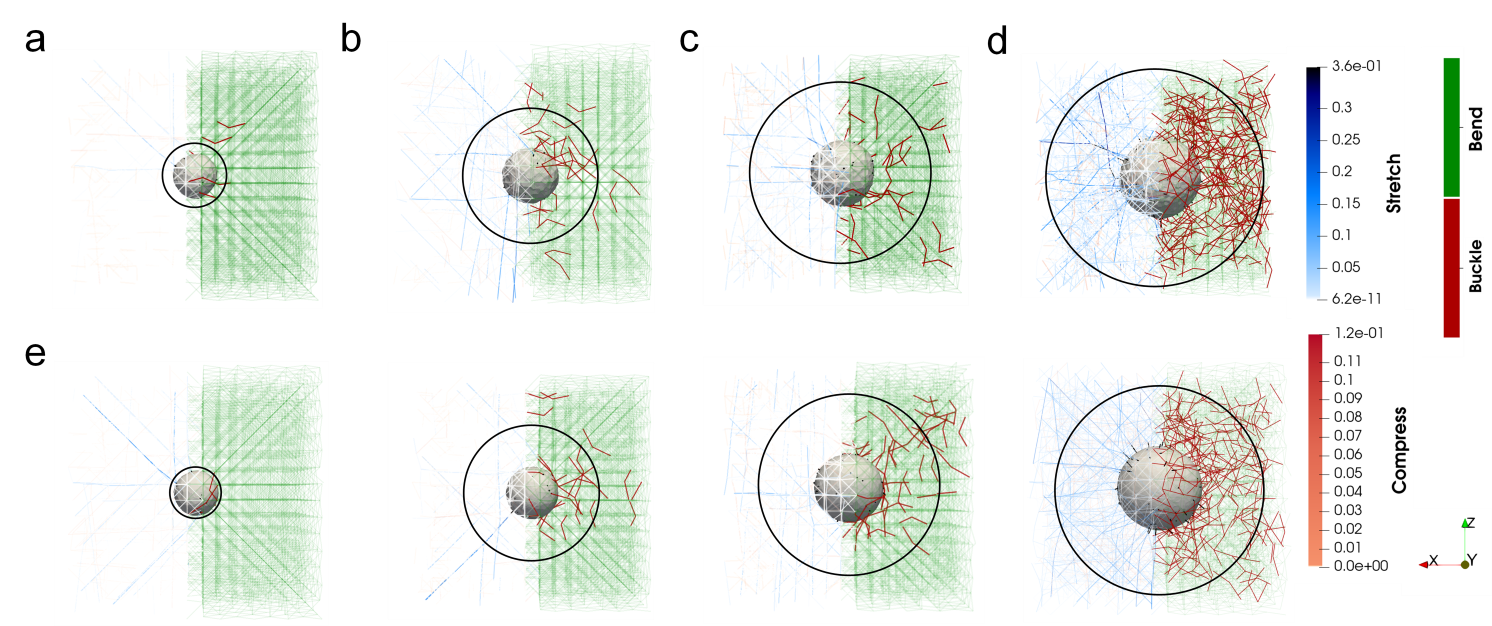}
\caption{Snapshots of fiber stress and buckle distribution for networks containing (a--d) a solid-like spheroid and (e--h) a fluid-like spheroid at 0\% strain, with radii 3, 3.78, 4.77, and 6, respectively. The left half of each box shows fiber stretching and compression forces, while the right half shows buckled or bent fibers. A bond is considered buckled if the angle of its constituent bonds is smaller than \(150^\circ\). The black circles contain 75\% buckled bonds, with radii 4.0, 8.7, 10.5, and 15.0 for the solid spheroid (a--d), and radii 3.4, 9.0, 10.7, and 14.2 for the fluid spheroid (e--h), respectively. Color intensity represents the magnitude of fiber tension, ranging from low to high.}
\label{fig:supp_pretension_snapshots_single}
\end{figure}

\begin{figure}[H]
\centering
\includegraphics[width=0.8\linewidth]{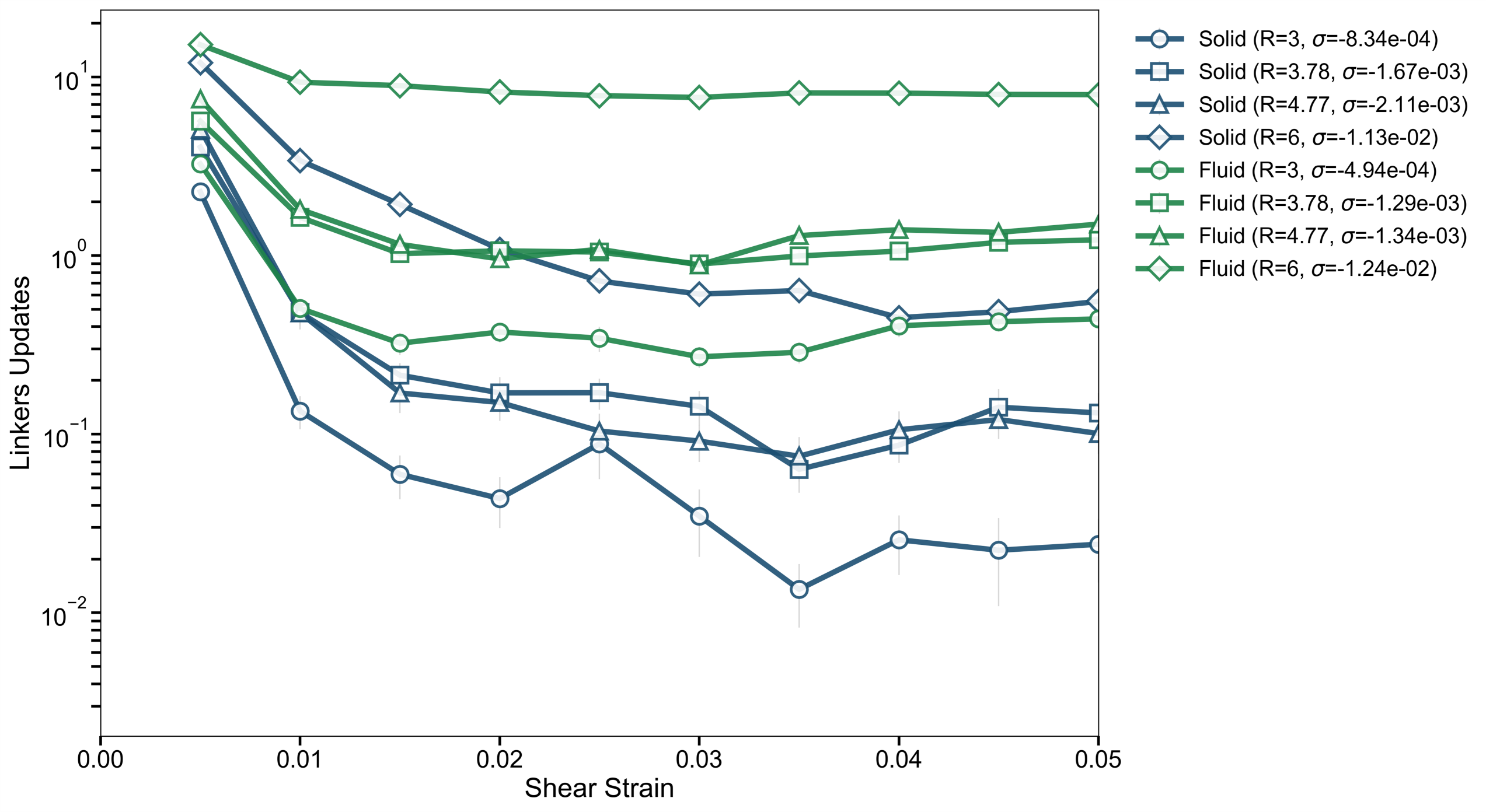}
\caption{Comparison of linker-update activity between solid-like and fluid-like spheroids as spheroid volume is increased. The update count measures newly assigned linkers following geometry- or topology-associated removal events. Fluid-like spheroids exhibit higher lability.}
\label{fig:supp_linker_lability_radius}
\end{figure}

\begin{figure}[H]
\centering
\includegraphics[width=0.8\linewidth]{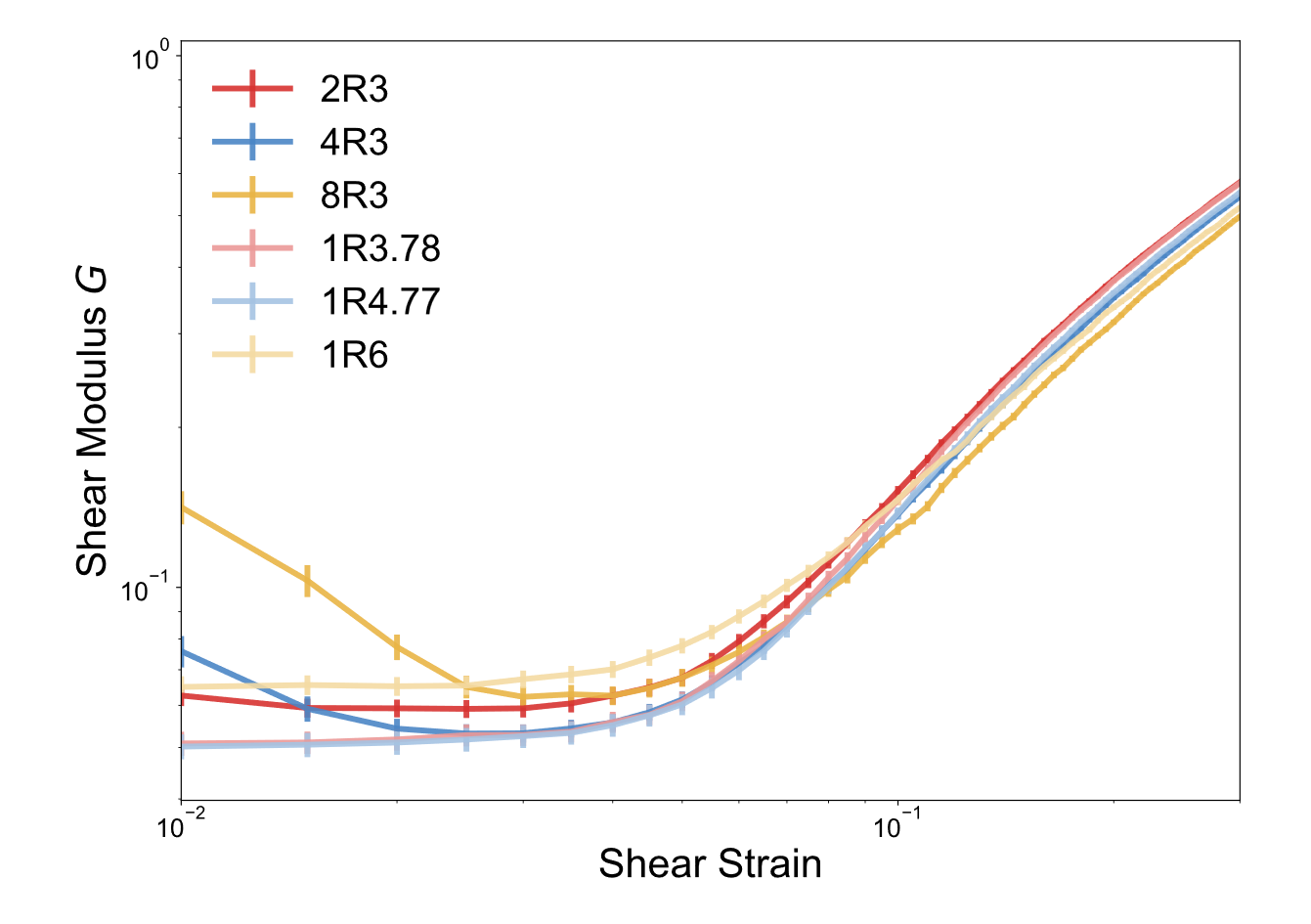}
\caption{Shear modulus as a function of shear strain for networks embedded with solid-like spheroids in aggregated and dispersed states, respectively. The notation 2R3, 4R3 and 8R3 means 2 spheroids, 4 spheroids and 8 spheroids with radius 3, respectively. The notation 1R3.78, 4R4.77 and 1R6 means 1 spheroid with radius 3.78, 4.77 and 6, respectively. Dispersing the volume into smaller inclusions decreases the inter-inclusion spacing \(\xi_p\) and increases the macroscopic modulus.}
\label{fig:supp_aggregated_vs_dispersed}
\end{figure}

\begin{figure}[H]
\centering
\includegraphics[width=1\linewidth]{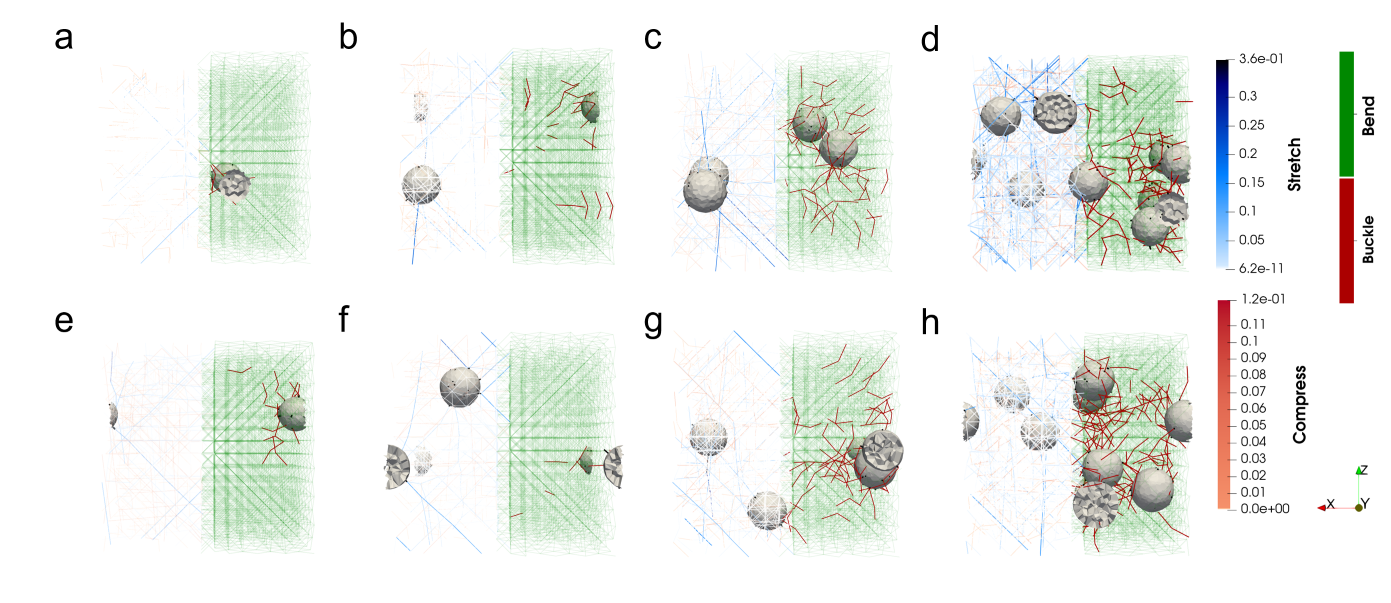}
\caption{Snapshots of fiber stress and buckle distribution for networks containing 1, 2, 4, and 8 (a--d) solid-like spheroids and (e--h) fluid-like spheroids at 0\% strain.  Left half box exhibits fibers stretch and compress force and right half box exhibits fibers buckle or bend. If angle of  constitute bonds smaller than \(150^0\), they are considered as buckle. Color intensity represents fiber tension magnitude, ranging from low to high. }
\label{fig:supp_pretension_snapshots}
\end{figure}

\begin{figure}[H]
\centering
\includegraphics[width=0.8\linewidth]{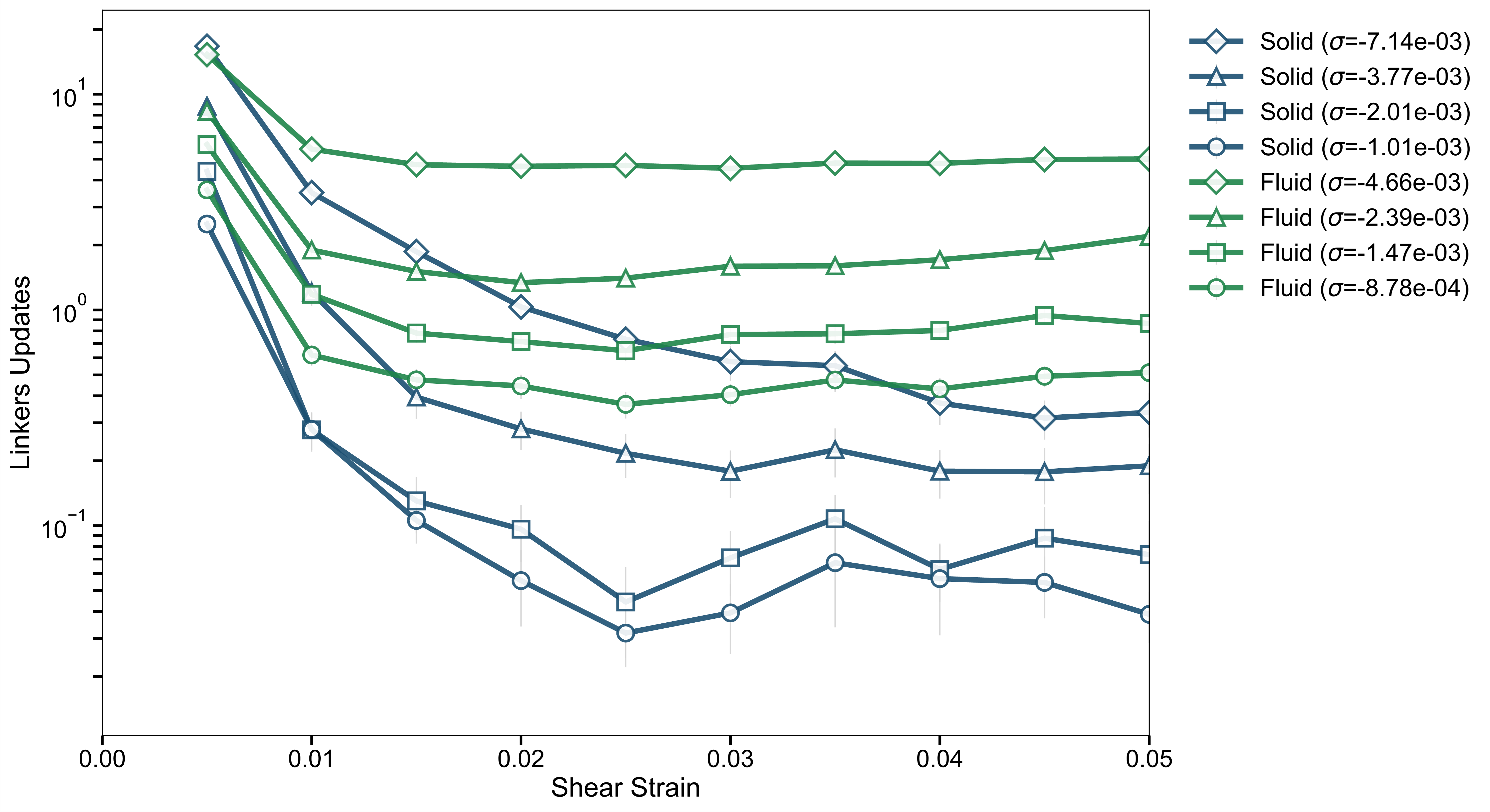}
\caption{Comparison of linker-update activity in systems containing multiple solid-like or fluid-like spheroids. The update count measures newly assigned linkers following geometry- or topology-associated removal events. Fluid-like spheroids exhibit higher lability.}
\label{fig:supp_linker_lability_number}
\end{figure}

\begin{figure}[H]
\centering
\includegraphics[width=\linewidth]{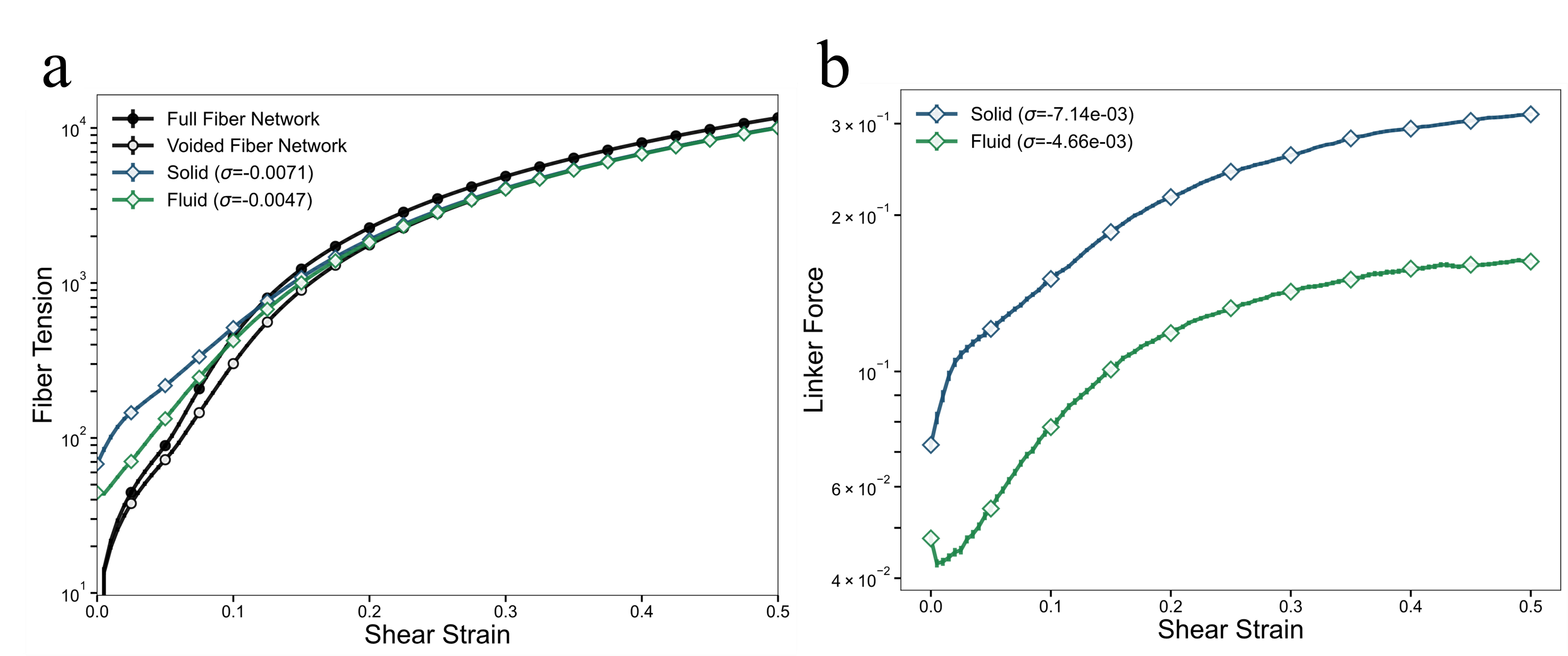}
\caption{\textbf{Large-strain evolution of fiber tension and linker force in spheroid--fiber composites.}
(a) Fiber tension as a function of applied shear strain $\gamma$ for an intact fiber network, a matched eight-void fiber network, and networks containing eight solid-like or fluid-like spheroids with active contractile linkers.
(b) Linker force as a function of $\gamma$ for the corresponding solid-like and fluid-like spheroid systems.
Each spheroid and corresponding cavity has radius $R=3$.
The eight-void reference contains the same cavity geometry as the spheroid-containing systems but no spheroids or spheroid--matrix linkers.
The values of $\sigma$ in the legends denote the fiber prestress at $\gamma=0$, measured after active contraction and mechanical relaxation but before shear; negative values correspond to contractile states.
The linker force remains finite throughout loading and differs between solid-like and fluid-like spheroids, even as their apparent moduli approach the voided-network response at large strain.
Each curve represents an average over $100$ independent simulations.}
\label{fig:large_strain_forces}
\end{figure}

\begin{figure}[H]
\centering
\includegraphics[width=1\linewidth]{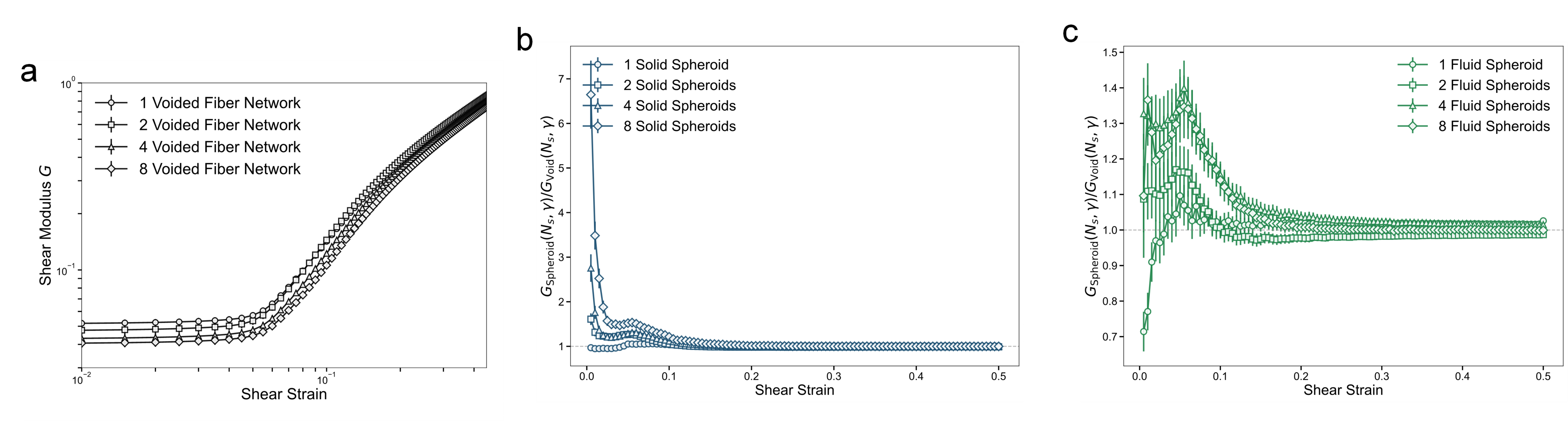}
\caption{\textbf{Comparison between spheroid-containing and corresponding voided fiber networks.} (a) Shear modulus $G$ as a function of shear strain $\gamma$ for fiber networks containing 1, 2, 4, or 8 spherical voids of radius $R=3$. (b,c) Modulus ratio $G_{\mathrm{spheroid}}/G_{\mathrm{void}}$ for networks containing (b) solid-like and (c) fluid-like spheroids. For each spheroid number $N_s$, the reference network contains the same number of spherical voids with the same radius. The approach of $G_{\mathrm{spheroid}}/G_{\mathrm{void}}$ toward unity demonstrates convergence to the corresponding voided-network response at large strain.}
\label{fig:supp_voided_fiber_network}
\end{figure}

\begin{figure}[H]
\centering
\includegraphics[width=0.6\linewidth]{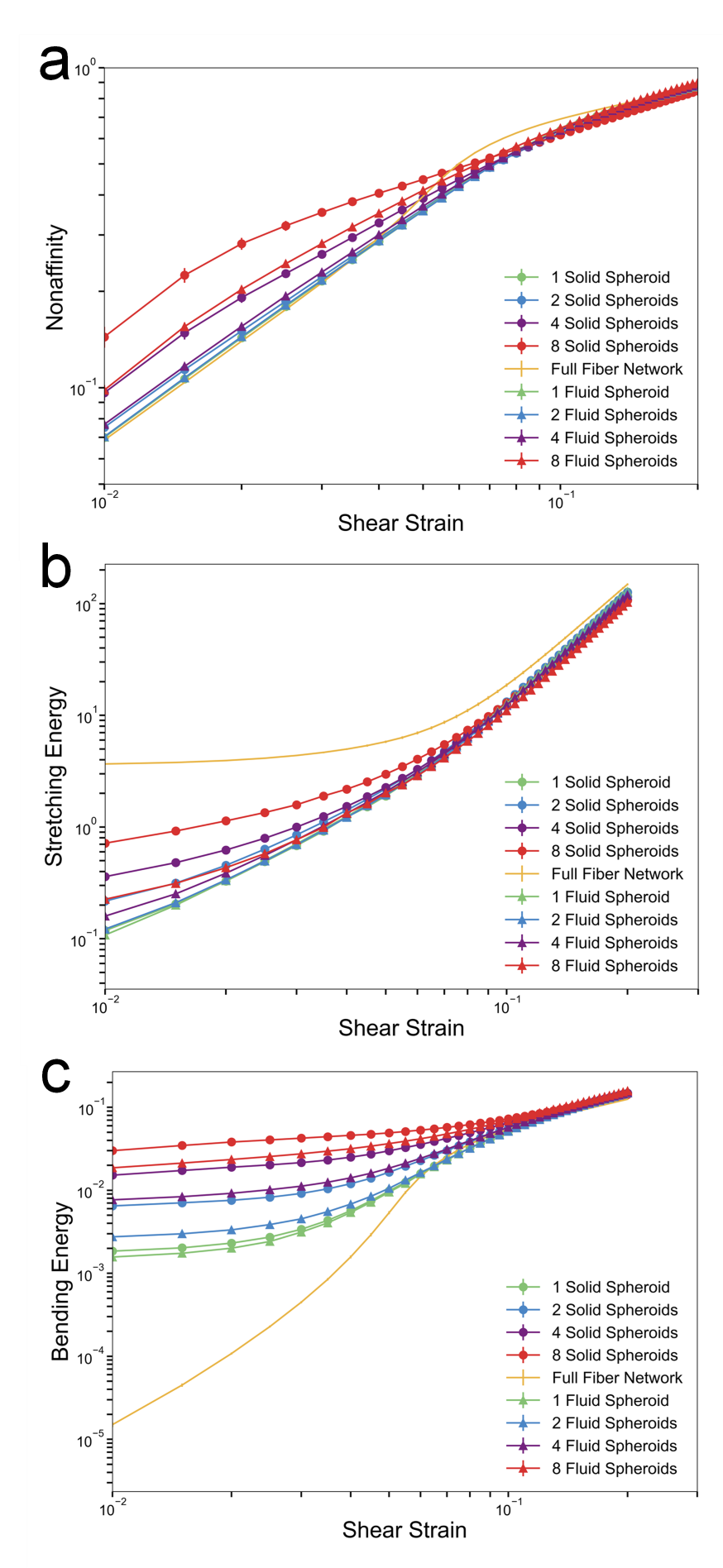}
\caption{\textbf{Nonaffinity and elastic-energy components in embedded fiber networks.}
(a) Global nonaffinity, (b) total fiber-stretching energy, and (c) total fiber-bending energy as functions of shear strain for the intact full fiber network and networks containing different numbers of solid-like or fluid-like spheroids. All spheroid-containing systems have radius $R=3$ and use $K_B=0.001$.}
\label{fig:supp_nonaffinity_energy_partition}
\end{figure}

\begin{figure*}[t]
\centering
\includegraphics[width=0.95\textwidth]{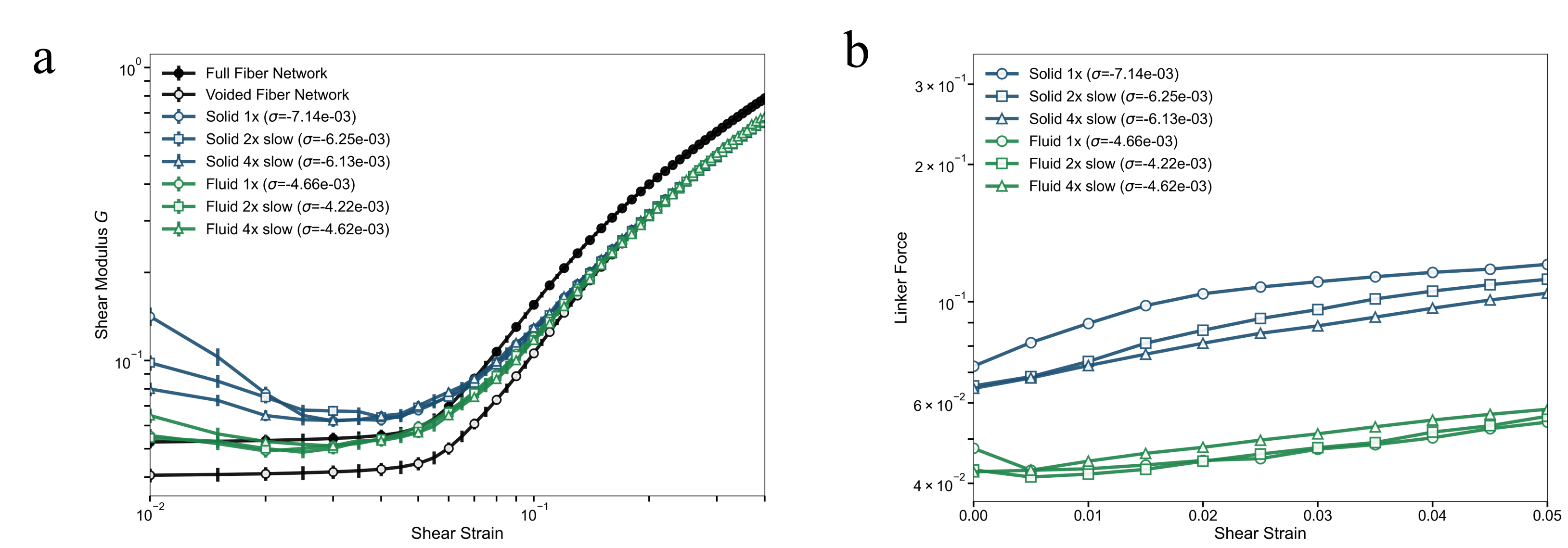}
\caption{
\textbf{Sensitivity of network mechanics to the linker contraction--relaxation protocol.}
(a) Apparent shear modulus $G$ versus shear strain $\gamma$ for the intact fiber network, the eight-void reference, and networks containing eight solid-like or fluid-like spheroids with $R=3$. The $1\times$ condition uses the standard linker rest-length update, whereas the $2\times$ slow and $4\times$ slow conditions reduce the rest-length decrement per FIRE iteration by factors of two and four, respectively, while retaining the same final rest length. (b) Corresponding linker force at small strains. Values of $\sigma$ in the legends denote the pre-shear fiber prestress after zero-strain contraction and relaxation.
}
\label{fig:SI_linker_contraction}
\end{figure*}

\end{document}